\documentclass[aps,prd,showpacs,superscriptaddress,nofootinbib,nobibnotes,twocolumn,linenumbers,10pt]{revtex4}

\usepackage{color,amsmath,amssymb,graphicx,latexsym,txfonts}
\usepackage{threeparttable,multirow,subfigure,booktabs,lineno}
\usepackage[colorlinks=true,linkcolor=blue,urlcolor=blue,
            citecolor=green,bookmarks=true,bookmarksnumbered=true,
            breaklinks=true,pdfpagemode=Fullscreen,pdfstartview=FitBH]{hyperref}

\def\jcap{J.\ Cosmol.\ Astropart.\ Phys.\ }

\begin{document}


\title{Measurement of the cosmic ray helium energy spectrum from 70 GeV 
to 80 TeV with the DAMPE space mission}

\author{F.~Alemanno}
\affiliation{Gran Sasso Science Institute (GSSI), Via Iacobucci 2, I-67100 L'Aquila, Italy}
\affiliation{Istituto Nazionale di Fisica Nucleare (INFN) -Laboratori Nazionali del Gran Sasso, I-67100 Assergi, L'Aquila, Italy}

\author{Q.~An}
\affiliation{State Key Laboratory of Particle Detection and Electronics, University of Science and Technology of China, Hefei 230026, China}
\affiliation{Department of Modern Physics, University of Science and Technology of China, Hefei 230026, China}

\author{P.~Azzarello}
\affiliation{Department of Nuclear and Particle Physics, University of Geneva, CH-1211, Switzerland}

\author{F.~C.~T.~Barbato}
\affiliation{Gran Sasso Science Institute (GSSI), Via Iacobucci 2, I-67100 L'Aquila, Italy}
\affiliation{Istituto Nazionale di Fisica Nucleare (INFN) -Laboratori Nazionali del Gran Sasso, I-67100 Assergi, L'Aquila, Italy}

\author{P.~Bernardini}
\affiliation{Dipartimento di Matematica e Fisica E. De Giorgi, Universit\`a del Salento, I-73100, Lecce, Italy}
\affiliation{Istituto Nazionale di Fisica Nucleare (INFN) - Sezione di Lecce, I-73100, Lecce, Italy}

\author{X.~J.~Bi}
\affiliation{Institute of High Energy Physics, Chinese Academy of Sciences, Yuquan Road 19B, Beijing 100049, China}
\affiliation{University of Chinese Academy of Sciences, Yuquan Road 19A, Beijing 100049, China}

\author{M.~S.~Cai}
\affiliation{Key Laboratory of Dark Matter and Space Astronomy, Purple Mountain Observatory, Chinese Academy of Sciences, Nanjing 210023, China}
\affiliation{School of Astronomy and Space Science, University of Science and Technology of China, Hefei 230026, China}

\author{E.~Catanzani}
\affiliation{Istituto Nazionale di Fisica Nucleare (INFN) - Sezione di Perugia, I-06123 Perugia, Italy}

\author{J.~Chang} 
\affiliation{Key Laboratory of Dark Matter and Space Astronomy, Purple Mountain Observatory, Chinese Academy of Sciences, Nanjing 210023, China}
\affiliation{School of Astronomy and Space Science, University of Science and Technology of China, Hefei 230026, China}

\author{D.~Y.~Chen}
\affiliation{University of Chinese Academy of Sciences, Yuquan Road 19A, Beijing 100049, China}
\affiliation{Key Laboratory of Dark Matter and Space Astronomy, Purple Mountain Observatory, Chinese Academy of Sciences, Nanjing 210023, China}

% ...

\collaboration{DAMPE Collaboration}\email{dampe@pmo.ac.cn}

\date{\today}

\begin{abstract}
The measurement of the energy spectrum of cosmic ray helium nuclei 
from 70 GeV to 80 TeV using 4.5 years of data recorded by the DArk 
Matter Particle Explorer (DAMPE) is reported in this work. 
A hardening of the spectrum is observed at an energy of about 
1.3 TeV, similar to previous observations. In addition, a spectral 
softening at about 34 TeV is revealed for the first time with large 
statistics and well controlled systematic uncertainties, with an 
overall significance of $4.3\sigma$. The DAMPE spectral measurements 
of both cosmic protons and helium nuclei suggest a particle charge 
dependent softening energy, although with current uncertainties a 
dependence on the number of nucleons cannot be ruled out.
\end{abstract}


\maketitle

{\it Introduction.} ---
Galactic cosmic rays (GCRs) are energetic particles traveling across 
the Galaxy as high-energy beams, and are a unique probe to explore the 
astrophysical particle accelerators and the interstellar medium of the 
Galaxy \cite{Grenier:2015egx}. The energy spectrum of GCRs is expected 
to be a power-law form for energies below the ``knee'' (at $3-4$ PeV) 
according to the canonical shock acceleration of particles. However, 
several experiments surprisingly observed changes in the power-law 
spectral indices $\gamma$ for protons, helium and heavy nuclei 
\cite{ATIC,CREAM,PAMELA,AMS,AMSHe,DAMPEp,CALET,NUCLEON}. 
Specifically, the spectra of GCRs become harder by $\Delta \gamma 
\simeq 0.1-0.2$ at kinetic energies (or rigidities) of several hundred 
GeV/n (or GV), and become softer again by $\Delta \gamma \simeq -0.3$ 
at energies of $15-30$ TeV (for protons and possibly helium). 
The deviations from single power-law of the spectra motivate extensive 
investigations for deeper understanding of the acceleration and propagation 
mechanisms or of new possible GCR sources (e.g., \cite{Model}).

Precise measurements of the GCR spectra, particularly for individual
species, are mainly from magnetic spectrometers such as the Payload for 
Antimatter-Matter Exploration and Light-nuclei Astrophysics (PAMELA)
and Alpha Magnetic Spectrometer (AMS-02) whose maximum measurable
rigidity can reach only few TV. Direct measurements at higher energies
were mostly done with balloon-borne calorimeter experiments in the
past decades, and the uncertainties (both statistical and systematic)
are somewhat large, hindering a good understanding of the spectral
features above TeV energies \cite{ATIC,CREAM,NUCLEON,SOKO,TRACER}.

The DArk Matter Particle Explorer (DAMPE; \cite{DAMPE}) is a 
satellite-borne particle and $\gamma$-ray detector launched on
December 17, 2015. It consists of a Plastic Scintillator Detector 
(PSD) for charge measurement \cite{PSD,PSDCharge}, 
a Silicon Tungsten tracKer-converter (STK) for trajectory
measurement \cite{STK,STKalign,Li:2019}, a Bi$_3$Ge$_4$O$_{12}$
electromagnetic calorimeter (BGO) for energy measurement and
electron-hadron discrimination \cite{BGO}, and a NeUtron Detector (NUD) 
for additional electron-hadron discrimination \cite{Huang:2020skz}.
DAMPE is expected to significantly improve the measurement precision 
of GCR spectra up to 100 TeV energies, due to its large 
acceptance and a good energy resolution ($\sim$$1.5\%$ for electrons 
and $\gamma$-rays \cite{DAMPEe} and $\sim$$30\%$ for nuclei \cite{DAMPEp}). 
Dedicated calibrations of each sub-detector show that the instrument
works very stably on-orbit \cite{DAMPECAL}.
In this letter we report the measurements of the helium spectrum 
with kinetic energies from 70 GeV to 80 TeV using 4.5 years of the 
DAMPE flight data. Our results give the first precise measurement 
of the helium spectral structure above TeV energies.

{\it Monte Carlo simulations.} ---
Extensive Monte Carlo (MC) simulations were carried out to explore 
the response to particles in the detector. The results presented in this
work are based on the GEANT4 toolkit of version 4.10.5 \cite{GEANT} 
with the FTFP$\_$BERT physics list for helium nuclei between $10$ GeV 
and $500$ TeV. For the higher energies ($>25$~TeV/n) we also tested 
the EPOS$\_$LHC model via linking the GEANT4 toolkit with the CRMC 
interface \cite{CRMC}, and found that the differences were negligible
($\lesssim1\%$). 
The test beam data at 40 GeV/n and 75 GeV/n were used to validate the 
simulation, and we found a good agreement between data and simulation 
\cite{testbeam}. The simulated events were generated with an isotropic 
source and an $E^{-1}$ spectrum. During the analysis, the simulation 
data were re-weighted to an $E^{-2.6}$ spectrum, and the systematic 
uncertainties from different spectral indices were studied. 
The isotope $^3$He was mixed with the $^4$He sample following the 
measurements of AMS-02 \cite{AMSHeIso}, with an extrapolation at 
higher energies. For protons we used the GEANT4 FTFP$\_$BERT physics 
list between 10 GeV and 100 TeV, and the DPMJET3 model via the 
CRMC-GEANT4 interface between 100 TeV and 1 PeV \cite{DAMPEp}. 

To evaluate the impact from the uncertainties of hadronic models, 
we also performed simulations with the FLUKA version 2011.2x 
\cite{FLUKA}, which uses DPMJET3 for nucleus-nucleus interaction 
above $5$ GeV/n. The same analysis procedures based on the two 
simulation samples were carried out, and the final differences of 
the energy spectra were taken as systematic uncertainties
from the hadronic models \cite{Jiang:2020nph}. 

{\it Event selection.} ---
In this analysis we used $54$ months of the flight data recorded by DAMPE 
from January 1st, 2016 to June 30th, 2020. The events when the detector
traveled across the South Atlantic Anomaly (SAA) region were excluded.
After subtracting the instrumental dead time, which is 3.0725 ms per
event ($\sim$$17.2\%$ of the operation time), the on-orbit calibration 
time ($\sim$$1.7\%$), the time between September 9, 2017 and September 13, 
2017 when a giant solar flare affected the operation status of the 
detector \cite{solarflare}, and the SAA passage time ($\sim$$4.9\%$), 
we got a total live time of $1.08\times10^8$~s, corresponding to 
$76.2\%$ of the total operation time. 

\begin{figure*}[!htb]
\centering
\includegraphics[width=0.33\textwidth]{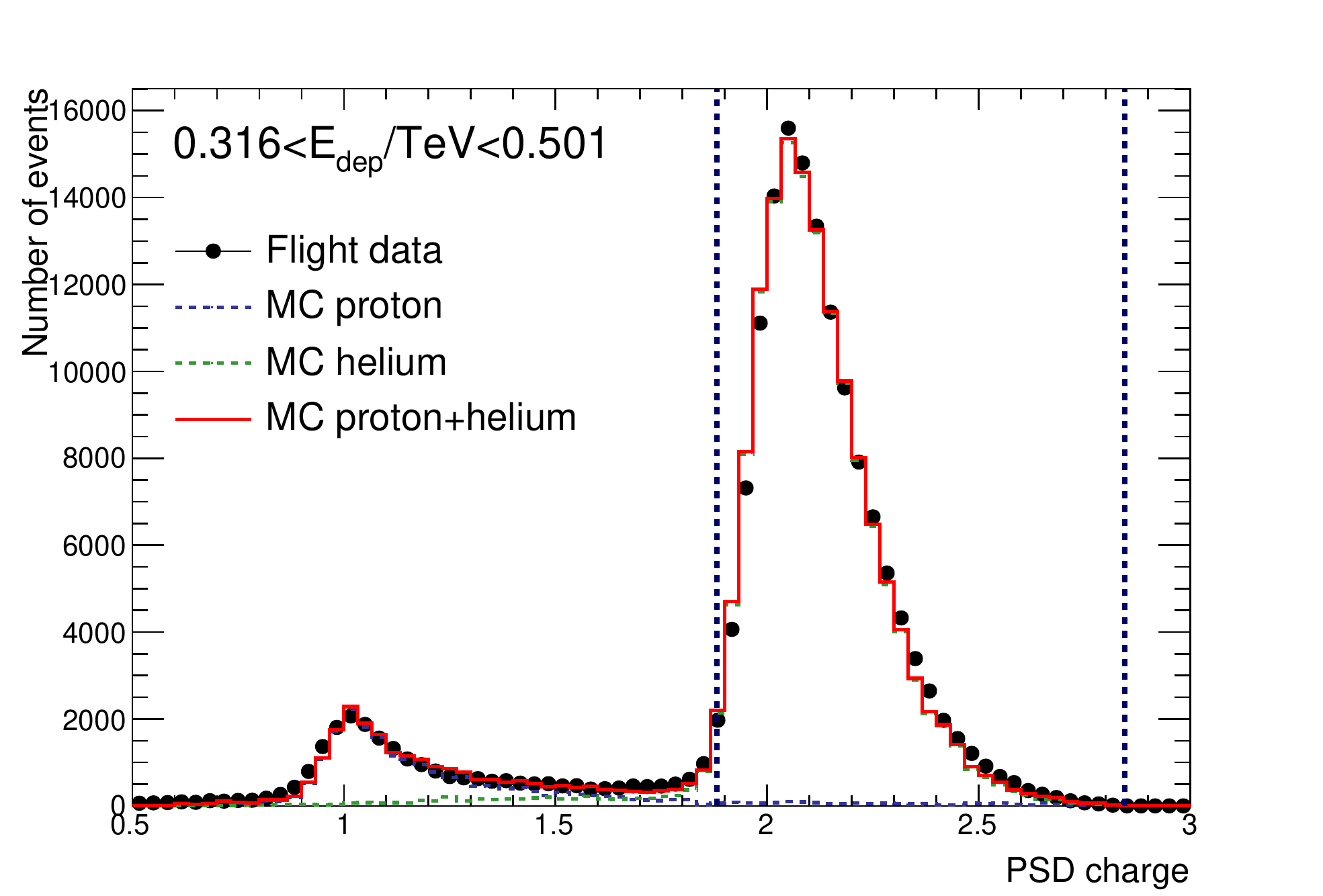}
\includegraphics[width=0.33\textwidth]{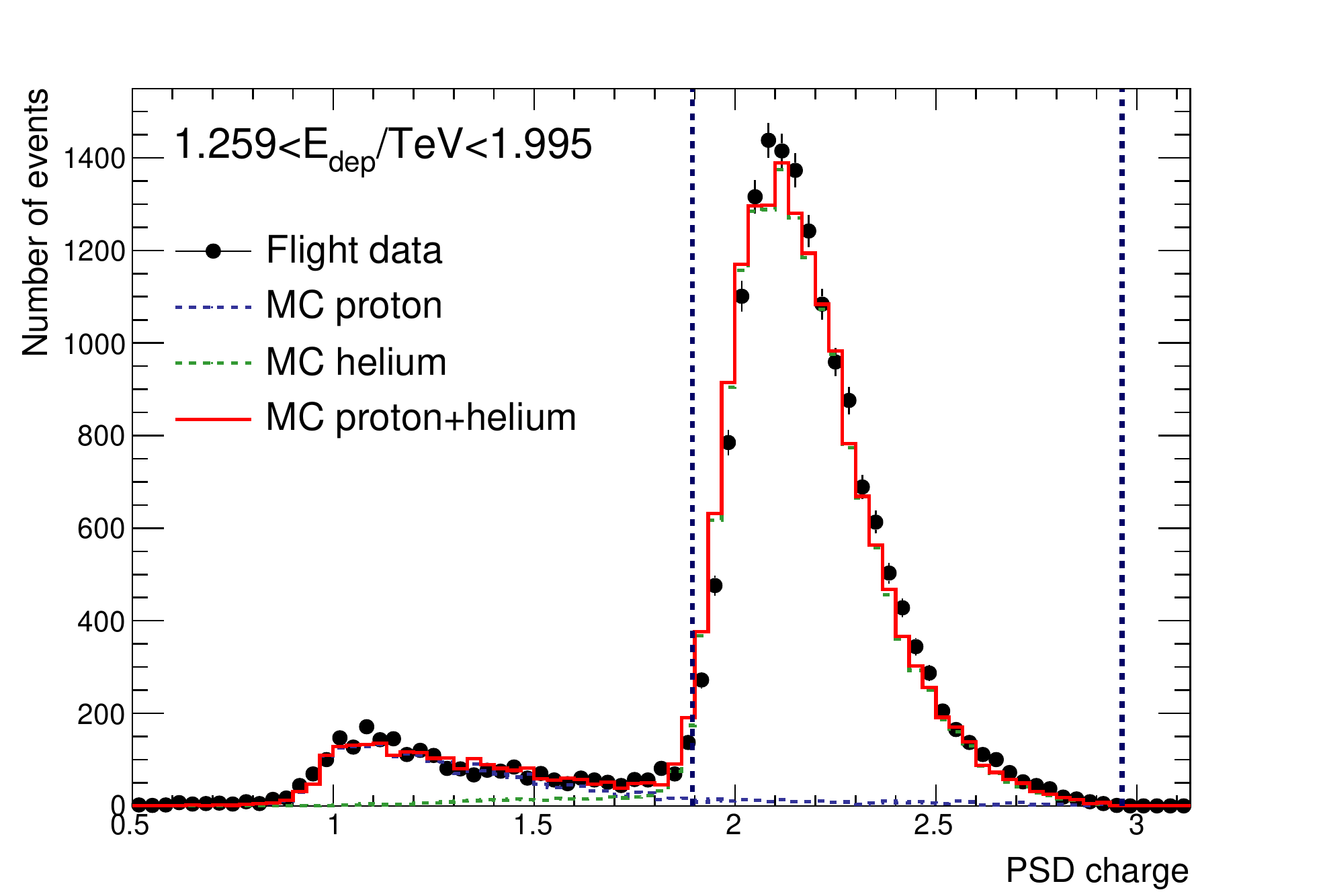}
\includegraphics[width=0.33\textwidth]{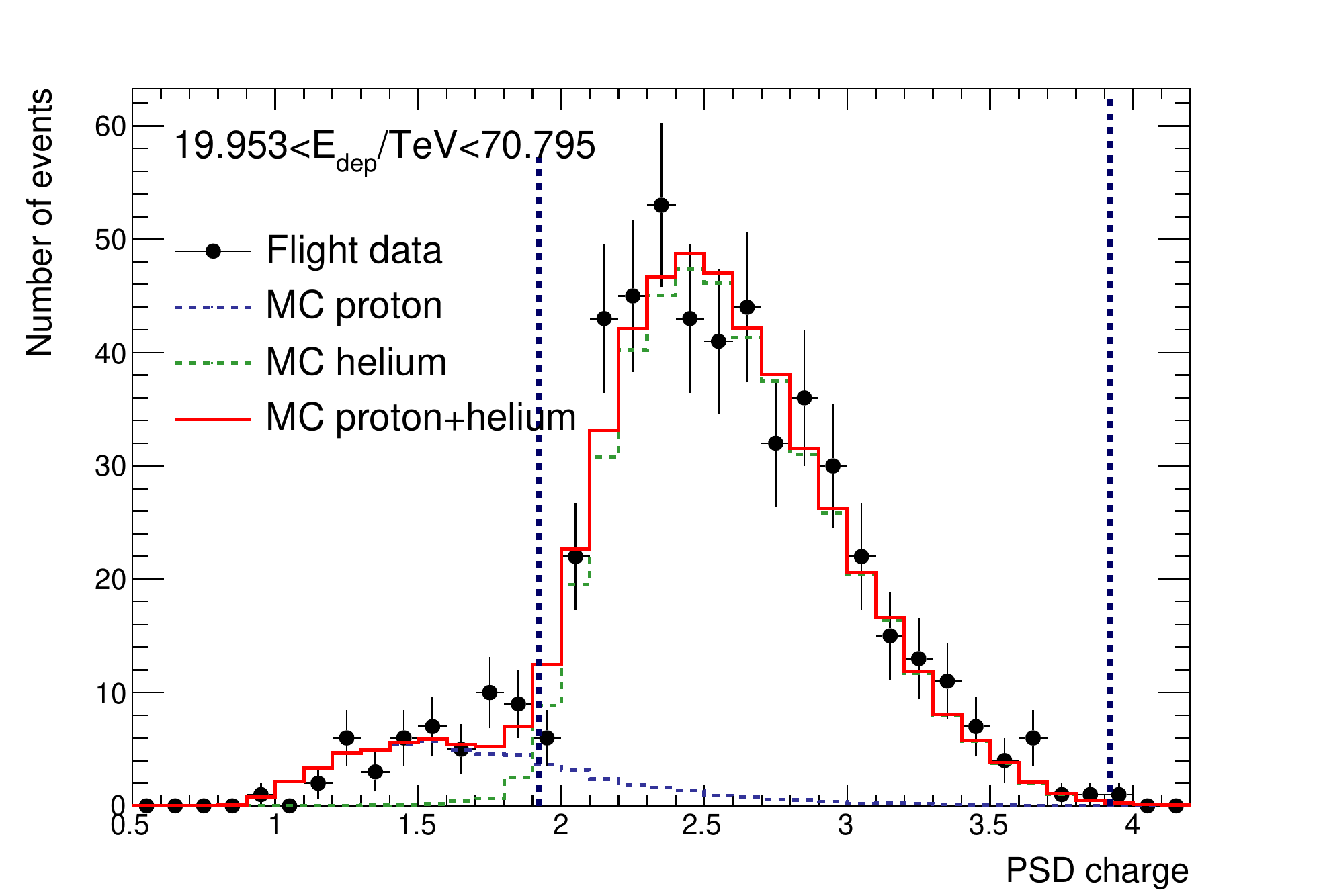}
\caption{The distributions of PSD charge, defined as the minimum charge 
value of the two PSD layers, for events with deposited energy ranges 
$316-501$ GeV (left), $1259-1995$ GeV (middle) and $19.95-70.79$ TeV 
(right). The flight data are shown in black points. The histograms 
show the distributions of the best-fit proton MC (blue), helium MC 
(green), and proton + helium MC (red). The vertical dashed lines 
indicate the PSD charge range used to select helium candidate events.}
\label{Fig:Templates}
\end{figure*}

The data were further filtered with the following steps.
\begin{itemize}
\item {\it Pre-selection.} 
A sample of good events was selected with a series of pre-selection 
criteria. The events passing the High Energy Trigger (HET) were used 
in this analysis. The HET requires that the energy depositions in 
the first three BGO layers are higher than about $13$ times the proton 
minimum ionizing particle (MIP) energy (about 23 MeV in one layer) 
and in the fourth layer is higher than 2.4 times proton MIP energy 
\cite{Zhang:2019}. Besides the HET, we further required that the 
energy deposition in the first two BGO layers was smaller than that in 
the third and fourth layers. These conditions guarantee that the shower
starts in the beginning of the calorimeter and results in a fairly good 
energy resolution ($\sim$$28\%$ at 1 TeV and $\sim$$34\%$ at 50 TeV). 
To avoid the geomagnetic rigidity cut-off effect \cite{GEOCUTOFF}, 
the energy deposition in the first 13 layers of the BGO calorimeter 
was required to be larger than 20 GeV. In this work, the first 13 
layers of the calorimeter were used to measure the event energy in order 
to minimize the effect of the saturation of readout electronic which is 
most severe in the last BGO layer due to the high gain of this layer. 
Finally, the energy recorded in each layer was required to be less 
than $35\%$ of the total deposited energy in the first 13 layers. 
This requirement effectively excludes particles entering from the 
sides of the detector.

\item {\it STK Track selection.}
The number of hits of the reconstructed tracks was required to be $\ge 3$. 
The track with the maximum total ADC was chosen if there were more than 
one candidate tracks passing the number of hits selection,  
and the reduced $\chi^2$ of the track fitting was required to be 
smaller than 35. Then we required a match between the selected STK track 
and the reconstructed BGO track, with the following two conditions: 
a) the projected distances on each PSD layer for the STK track and the 
BGO track were smaller than 90 mm, and b) the average projected distances 
between the STK track and the centroids of the energy depositions in 
the first four BGO layers were smaller than 25 mm. Furthermore, to 
ensure a good shower containment, the reconstructed track was required 
to be fully contained in the PSD, STK and BGO sub-detectors, and the 
bar with the maximum energy deposition in each layer was required
to be not at the edge of the calorimeter.

\item {\it Charge selection.}

The helium candidates were selected by the charge measured in PSD and STK. 
The signal of the first hit in the STK track was requested to be higher 
than 2.5 times of the MIP-equivalent signal. This is a very loose STK 
charge selection to suppress proton events. To properly account for the 
increase of the energy deposition in the PSD bars with higher particle
energies (due to the Bethe-Bloch formula and the backscattering particles), 
a deposited-energy-dependent selection of the charge reconstructed in 
both PSD layers ($Y$-layer for the first and $X$-layer for the second),
\begin{eqnarray}
&1.85+0.02\cdot\log{\frac{E_{\rm dep}}{10~{\rm GeV}}}& < Z_{X(Y)} < \nonumber\\
&2.8+0.007\cdot\left(\log{\frac{E_{\rm dep}}{10~{\rm GeV}}}\right)^{4.0}&,
\end{eqnarray} 
was adopted. Note that the energy-dependence was not considered in the 
PSD charge reconstruction \cite{PSDCharge} algorithm, and the 
``PSD charge'' here was not equivalent to the real particle charge.
Finally, the PSD charge reconstructed based on the selected track for 
both layers was required to be within a factor of 2.

Fig.~\ref{Fig:Templates} shows the PSD charge (the minimum of $X$ and $Y$
layer measurements\footnote{Note that we selected events using both 
charge in the $X$ and $Y$ layers of PSD. However, for the background 
estimate, the template fitting algorithm which will be described below
was applied to the one-dimensional PSD charge distribution defined as 
the minimum of $Z_X$ and $Z_Y$.}) 
distributions for three selected deposited energy bins, $316-501$ GeV, 
$1259-1995$ GeV and $19.95-70.79$ TeV. The vertical dashed lines show 
the PSD charge selection conditions of Eq.~(1). After the STK first-point 
cut, proton candidates were heavily excluded, which enabled a pure 
helium sample to be selected in our analysis.

\end{itemize}

\begin{figure}[!htb]
\centering
\includegraphics[scale=0.45]{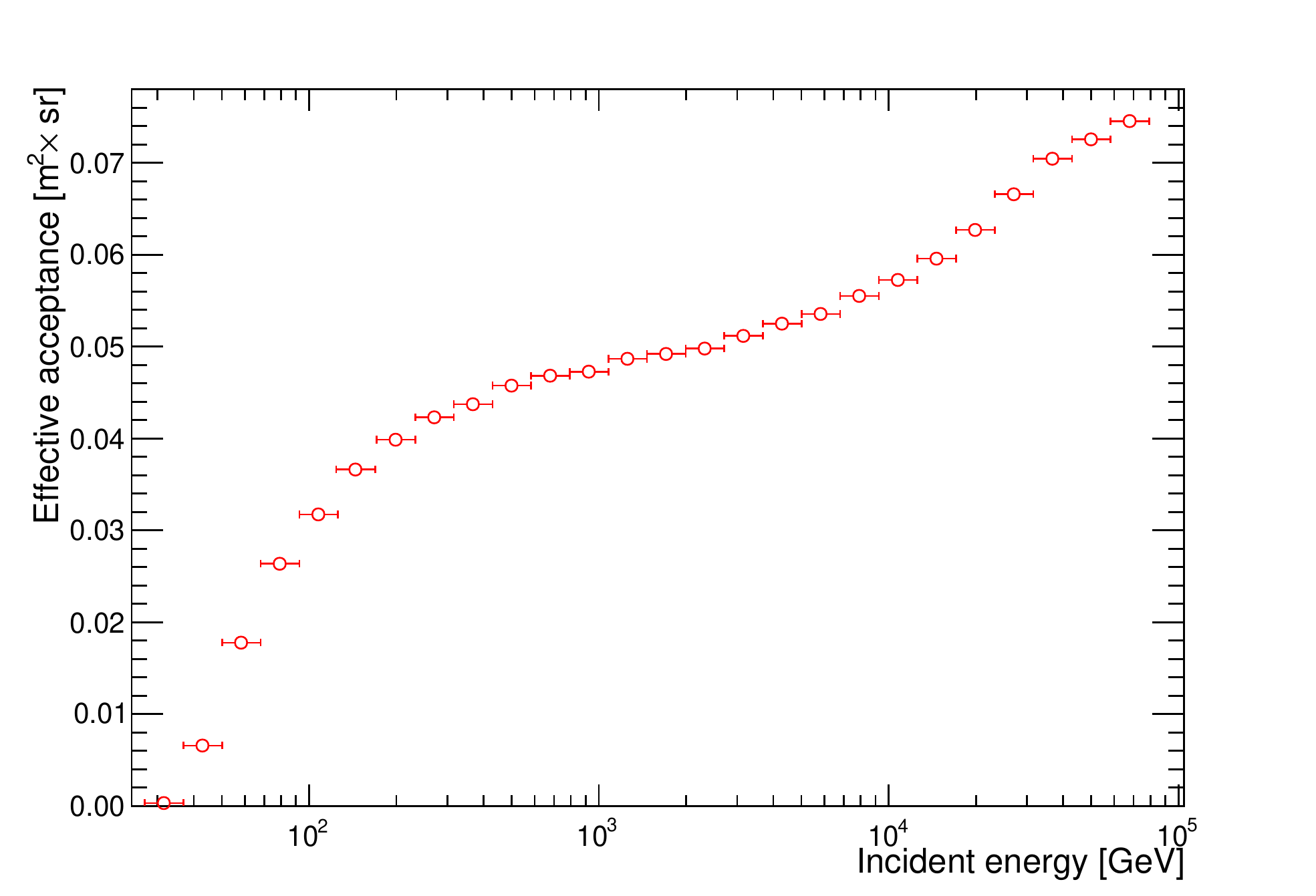}
\caption{Effective acceptance after all the analysis selections, 
as derived from the helium MC sample.}
\label{Fig:acceptance}
\end{figure}

The efficiencies of the selections were obtained from MC simulations.
The efficiencies vary with energy, and are about $42\%$, $84\%$, and 
$60\%$ for the pre-selection, track, and charge selections respectively,
at 1 TeV. For the validations of the main efficiencies one can refer to 
the {\tt Supplemental Material}. The effective acceptance after the 
selection, as a function of the incident energy for incoming helium 
nuclei, is shown in Fig.~\ref{Fig:acceptance}. Here the acceptance
in the $i$-th incident energy bin is computed as
\begin{equation}
{A}_{\mathrm{eff},i}=A_{\rm gen}\times \frac{N_{\mathrm{pass},i}}
{N_{\mathrm{gen},i}},
\label{eq:acceptance}
\end{equation}
where $A_\mathrm{gen}$ is the geometrical factor of the MC event generator 
sphere, $N_{\mathrm{pass},i}$ refers to the number of events passing the 
helium selection, and $N_{\mathrm{gen},i}$ is total number of generated events.
Noteworthy that the effective acceptance in this analysis is higher than
that of the proton analysis \cite{DAMPEp}, mainly due to the fact that 
helium events have a higher HET efficiency.

{\it Background subtraction.} ---
The main background in helium selection comes from protons. The Landau 
tail of the proton PSD charge distribution can extend readily to the 
helium PSD charge window. We employed the MC PSD charge distributions
as templates to fit to the data and estimated the background. 
The template fit was done on the one-dimensional PSD charge
distribution of the minimum of $Z_X$ and $Z_Y$. The PSD charge 
values reconstructed from the MC data and the flight data did not 
match precisely, especially at high energies, probably due to the 
backscattering particles which were not well modeled in the MC 
simulations. Therefore a smearing of the PSD charge distribution of 
the MC simulations was applied. The MC templates were shifted and 
stretched in each deposited energy bin to match with the peaks 
and widths of the flight data distributions for protons and helium 
nuclei individually. After the charge smearing, the MC results
can well fit the flight data, as shown in Fig.~\ref{Fig:Templates}. 
The contaminations of protons were then estimated by counting the
number of proton MC events lying in the helium charge window. 
The proton background varies between $\sim$$0.05\%$ for deposited 
energy of 20 GeV and $\sim$$4\%$ for 60 TeV. The background fraction 
in the full energy range is shown in Fig.~S5 of the 
{\tt Supplemental Material}.

{\it Energy measurements and spectral unfolding.} ---
In this work we used the first 13 layers of the BGO calorimeter to measure 
the energy of an event. We also performed two corrections of the energy 
measurement, as described below. A large energy deposit (approximately 
above 4 TeV) in a single BGO bar might result in a saturation even of
the low-gain readout channel \cite{DAMPE,SATURATION}. In most cases, 
the saturation occurred only for a single BGO bar per event. 
The adoption of 13 layers can effectively exclude the events with 
multiple saturated bars. We developed a method based on the MC simulations 
to correct the energy measurements for the saturated events \cite{SATURATION}. 
The other correction was performed in order to account for the Birks' 
quenching in BGO, which occurred for very low velocity secondary 
particles \cite{Birks:1951}. The effect is more significant for heavy 
nuclei since more secondary particles with large charge and low velocity 
are produced. We took this effect into account through adding a 
quenching term in the MC simulations when the ionization energy 
density was larger than 10 MeV/mm \cite{Wei:2020}. The quenching effect 
would result in $\sim$$2\%$ lower deposition of the shower energy for
$\sim$$80$ GeV incident energy, which translates into $\sim$$5.5\%$ 
higher helium flux at such an energy after the unfolding. An impact 
of the quenching effect at different energies is demonstrated in 
Fig.~S6 of the {\tt Supplemental Material}.

An unfolding procedure is necessary to obtain the incident energy 
spectrum, since only a fraction of the energy of a nucleus can be
deposited in the calorimeter. The {\tt observed} number of events 
$N_{{\rm obs},i}$ in the $i$-th deposited energy bin is related to 
the {\tt incident} numbers of events $N_{\rm inc}$ as
\begin{equation}
N_{{\rm obs},i}=\sum_{j} M_{ij} N_{{\rm inc},j},
\end{equation}
where $M_{ij}$ is the probability that particles in the $j$-th incident
energy bin contributing to the $i$-th deposited energy bin. The response 
matrix $M$ for helium nuclei from the GEANT4 FTFP\_BERT simulations
is given in Fig.~S7 of the {\tt Supplemental Material}.
In this work we used the Bayesian unfolding method \cite{DAGOSTINI} 
to derive the {\tt incident} numbers of events, which were then used 
to obtain the incident energy spectrum.

\begin{figure}[!htb]
\includegraphics[scale=0.45]{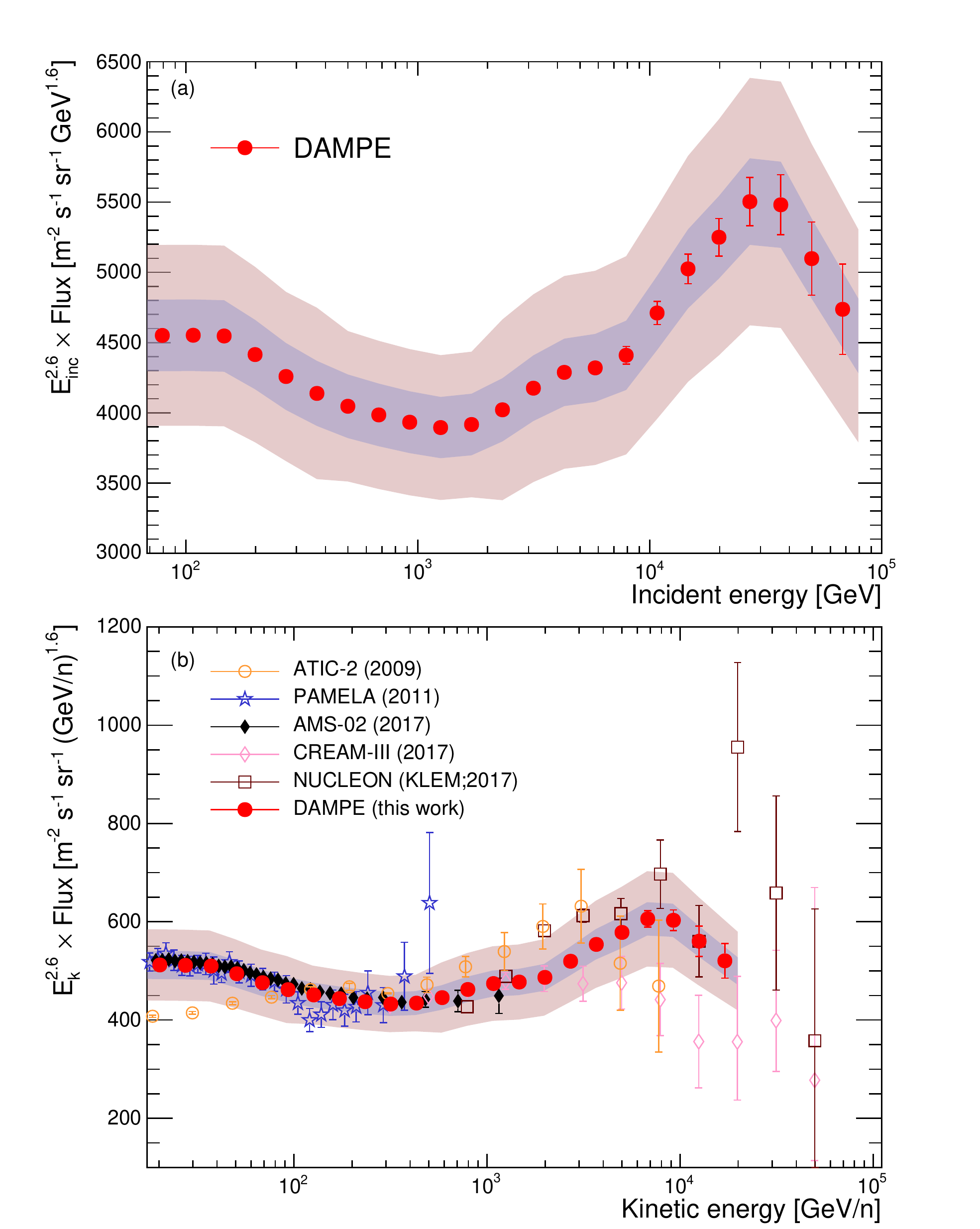}
\caption{Helium spectrum weighted by $E^{2.6}$ (top panel) measured by DAMPE. 
In the bottom panel, we compare the DAMPE spectrum (converted to kinetic
energy per nucleon assuming the AMS-02 measured $^3$He/$^4$He isotope
ratio \cite{AMSHeIso}) with previous measurements by PAMELA \cite{PAMELA}, 
AMS-02 \cite{AMSHe}, CREAM-III \cite{CREAM}, ATIC-2 \cite{ATIC}, and 
NUCLEON (KLEM) \cite{NUCLEON}. 
Error bars of the DAMPE data show the statistical uncertainties. 
The inner and outer shaded bands denote the systematic uncertainties 
from the analysis ($\sigma_{\rm ana}$) and the total systematic 
uncertainties including those from hadronic models 
$\left(\sqrt{\sigma_{\rm ana}^2+\sigma_{\rm had}^2}\right)$. 
For the PAMELA and AMS-02 results, the error bars contain both the 
statistical and systematic uncertainties added in quadrature. For the 
other measurements, only the statistical uncertainties are shown.}
\label{Fig:flux}
\end{figure}

{\it Results.} ---
The differential helium flux in the incident energy bin 
$[E_i,E_i+\Delta E_i]$ is given by 
\begin{equation}
\Phi(E_{i},E_{i}+\Delta E_{i})= \frac{N_{{\rm inc},i}}
{\Delta E_i~A_{{\rm eff},i}~\Delta t},
\end{equation}
where $\Delta E_i$ is the energy bin width, $N_{{\rm inc},i}$ is the 
unfolded number of events, $A_{{\rm eff},i}$ is the effective acceptance, 
and $\Delta t$ is the total live time. The helium spectrum weighted
by $E^{2.6}$ in the energy range from $70$ GeV to $80$ TeV is shown in 
the top panel of Fig.~\ref{Fig:flux}. The error bars show the statistical 
uncertainties, and the inner and outer shaded bands show the systematic 
uncertainties from the analysis procedure and the hadronic models, 
respectively. We also give the fluxes and the associated uncertainties 
of our measurement in Table S2 of the {\tt Supplemental Material}. 
The bottom panel of Fig.~\ref{Fig:flux} shows a comparison of
the DAMPE measurement with previous direct measurements by space and 
balloon-borne detectors \cite{PAMELA,AMSHe,CREAM,ATIC,NUCLEON}. 
Note that to convert the energy of a helium nucleus to the kinetic 
energy per nucleon, we assumed an isotope ratio of $^3$He/$^4$He from 
the AMS-02 measurements \cite{AMSHeIso}. For the results from other 
experiments, a pure $^4$He sample was usually assumed.

The statistical uncertainties come from the Poisson fluctuations of 
the number of detected events as well as the MC sample size. 
Due to the unfolding procedure, the statistical uncertainties 
cannot be simply translated into the incident energy bins. 
Following Ref.~\cite{DAMPEp}, we generated toy-MC samples based 
on the numbers of detected events and selected MC events following 
Poisson distributions in each deposited energy bin, and carried out the 
spectral unfolding for each simulated observation. The root-mean-square 
of the final helium fluxes in each incident energy bin is adopted as 
the statistical uncertainty.

There are several sources of systematic uncertainties of the measurements.
For the event selections, we used the differences between the flight
data and the MC simulations for control samples to evaluate the 
systematic uncertainties. The results turn out to be about $\sim$$4\%$ 
for the HET efficiency ($\sigma_{\rm HET}$), $\sim$$0.5\%$ for the track 
selection efficiency ($\sigma_{\rm track}$), $\sim$$3.5\%$ for the charge 
selection efficiency ($\sigma_{\rm charge}$). 
We re-weighted the spectrum of the MC simulations with spectral index
changing from 2.0 to 3.0, and found that the helium fluxes changed by
$\lesssim 1\%$. The analysis using energy measurements with 14 
layers of the BGO calorimeter led to $\lesssim1\%$ differences from 
the results presented here. These two were combined together to give
systematic uncertainties from the spectral unfolding, $\sigma_{\rm unf}$.
The $^3$He/$^4$He isotope ratio, which mainly affects the calculation
of the average number of nucleons, was estimated to contribute to
about $0.2\%$ ($\sigma_{\rm iso}$) of the fluxes at low energies 
($\sim$$100$ GeV) and even smaller at higher energies via varying 
the ratio by $\pm5\%$ which is conservative according to the AMS-02 
measurements \cite{AMSHeIso}. We also estimated the effect of background 
subtraction through varying the PSD charge selection of Eq.~(1) by 
$\pm5\%$, and found that the results differed by about $1\%-1.5\%$ 
($\sigma_{\rm bkg}$). The total systematic uncertainty from the 
analysis was given by the quadrature sum of the above uncertainties, 
which was about $5.6\%$. The absolute energy scale of the 
measurement, whose uncertainty was estimated to be $\sim$$1.3\%$ based 
on the geomagnetic cutoff of $e^{\pm}$ \cite{scale}, would result in 
a global but tiny shift of the spectrum, and was not included in the 
total systematic uncertainty. Different analyses obtained consistent 
results within the uncertainties.

The largest systematic uncertainty comes from the hadronic interaction
models. In this work we used the differences between the results based
on the GEANT4 and FLUKA simulations as the hadronic model systematic 
uncertainties, which turned out to be about $12\%-15\%$ for incident 
energies above $300$ GeV. At lower energies, we used the test beam data 
of Helium with kinetic energies 40 GeV/n and 75 GeV/n \cite{testbeam} 
to estimate the efficiencies and energy deposit ratios, and obtained 
the flux differences between the test beam data and simulation data 
of $\sim$$13\%$. Thus the systematic uncertainties from the hadronic 
model below 300 GeV were estimated as $13\%$. The statistical and 
systematic uncertainties for different incident energies are 
summarized in Fig.~S8 of the {\tt Supplemental Material}.

From Fig.~\ref{Fig:flux} we can observe that the Helium spectrum
experiences a hardening at $\sim$TeV energies and then shows a 
softening around $\sim$$30$ TeV. The spectral fitting (see the 
{\tt Supplemental Material} which includes Ref.~\cite{Abdollahi:2017nat}) 
gave a significance of the hardening of $24.6\sigma$, and a hardening 
energy of $(1.25^{+0.15}_{-0.12})$ TeV. 
What is more interesting is the softening feature which is clearly
shown in the DAMPE spectrum. A possible softening of the spectrum 
was reported by previous measurements \cite{CREAM,NUCLEON},
but the limited statistics and the large systematic uncertainties 
prevented a conclusion on this specific point. The significance of 
the softening from the DAMPE measurements is about $4.3\sigma$. 
The softening energy is found to be $34.4^{+6.7}_{-9.8}$ TeV, with 
a spectral change $\Delta\gamma=-0.51^{+0.18}_{-0.20}$. Together with 
the softening energy of the DAMPE proton spectrum, $13.6^{+4.1}_{-4.8}$ 
TeV \cite{DAMPEp}, the results are consistent with a charge-dependent 
softening energy of protons and helium nuclei, although a mass-dependent 
softening cannot be excluded by current data.

{\it Summary.} ---
The GCR helium spectrum from 70 GeV to 80 TeV is measured with 4.5 
years of the DAMPE data. We confirm the hardening feature of the 
helium spectrum reported by previous experiments. The hardening is 
smooth with a hardening energy of $\sim$$1.3$ TeV. The DAMPE data 
further reveals a softening feature at $\sim$$34$ TeV with a high 
significance of $4.3\sigma$. Combined with the proton 
spectrum, the softening energy is well consistent with a dependence 
on particle charge, although a dependence on particle mass can not 
be ruled out yet. These results will provide important implications 
in understanding GCR acceleration or propagation processes. 
Extending the DAMPE measurements to even higher energies is possible 
with new data and improved analysis performance.

{\it Acknowledgements.} ---
The DAMPE mission was funded by the strategic priority science and technology 
projects in space science of Chinese Academy of Sciences. In China the data 
analysis is supported by the National Key Research and Development Program 
of China (No. 2016YFA0400200), the National Natural Science Foundation of 
China (Nos. 11921003, 11622327, 11722328, 11851305, U1738205, U1738206, 
U1738207, U1738208, U1738127), the strategic priority science and 
technology projects of Chinese Academy of Sciences (No. XDA15051100), 
the 100 Talents Program of Chinese Academy of Sciences, the Young Elite 
Scientists Sponsorship Program by CAST (No. YESS20160196), 
and the Program for Innovative Talents and Entrepreneur in Jiangsu. 
In Europe the activities and data analysis are supported by the Swiss 
National Science Foundation (SNSF), Switzerland, the National Institute 
for Nuclear Physics (INFN), Italy, and the European Research Council (ERC) 
under the European Union's Horizon 2020 research and innovation programme 
(No. 851103).


\clearpage

\setcounter{figure}{0}
\renewcommand\thefigure{S\arabic{figure}}
\setcounter{table}{0}
\renewcommand\thetable{S\arabic{table}}

\section{Supplemental Material}

\subsection{Event topology}

\begin{figure}[!htb]
\centering
\includegraphics[width=0.48\textwidth]{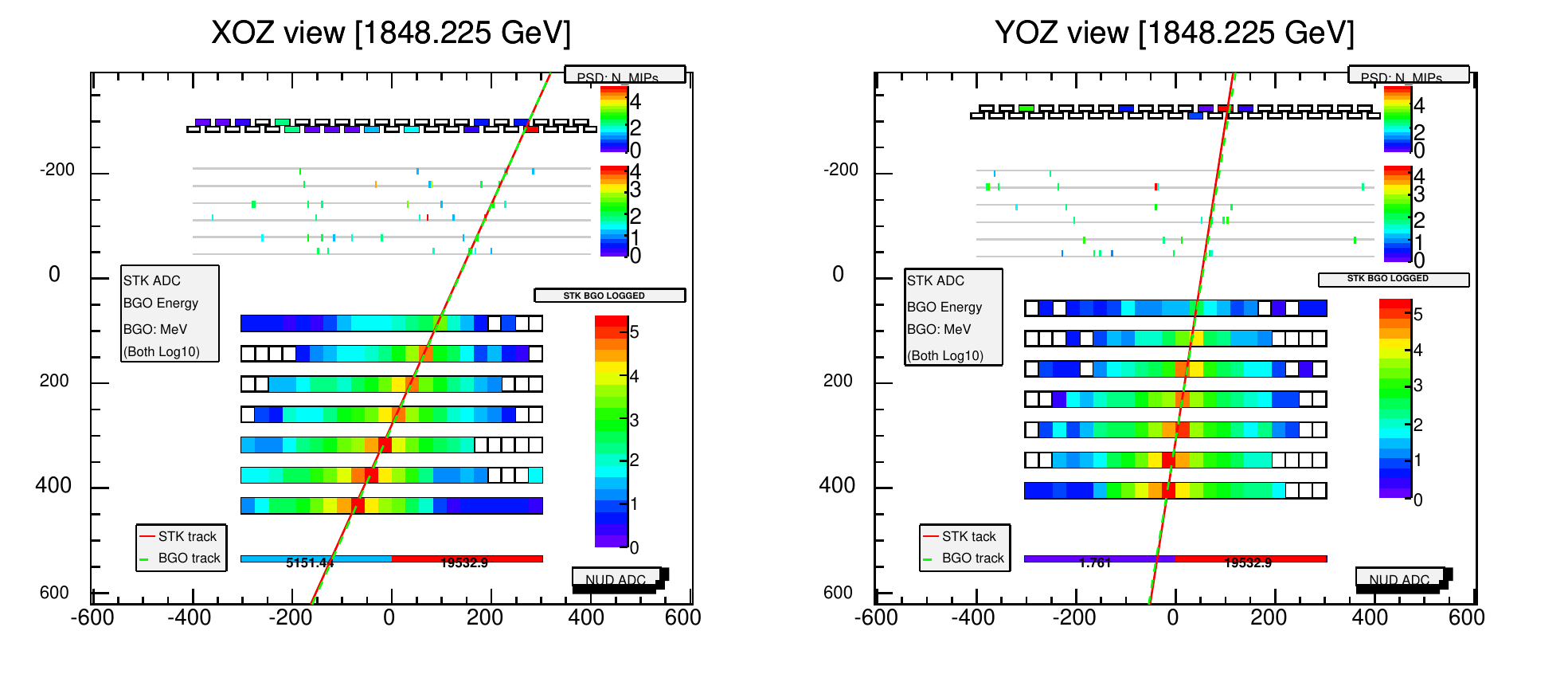}
\caption{Illustration of signals in all sub-detectors of DAMPE,
for a candidate helium event with a raw energy of $\sim$1.8 TeV.}
\label{Fig:Event}
\end{figure}

Fig.~\ref{Fig:Event} shows the topology of a typical event in 
all sub-detectors of DAMPE passing the helium selection.

\subsection{Efficiency validations}

\subsubsection{HET efficiency}

\begin{figure}[!htb]
\centering
\includegraphics[width=0.43\textwidth]{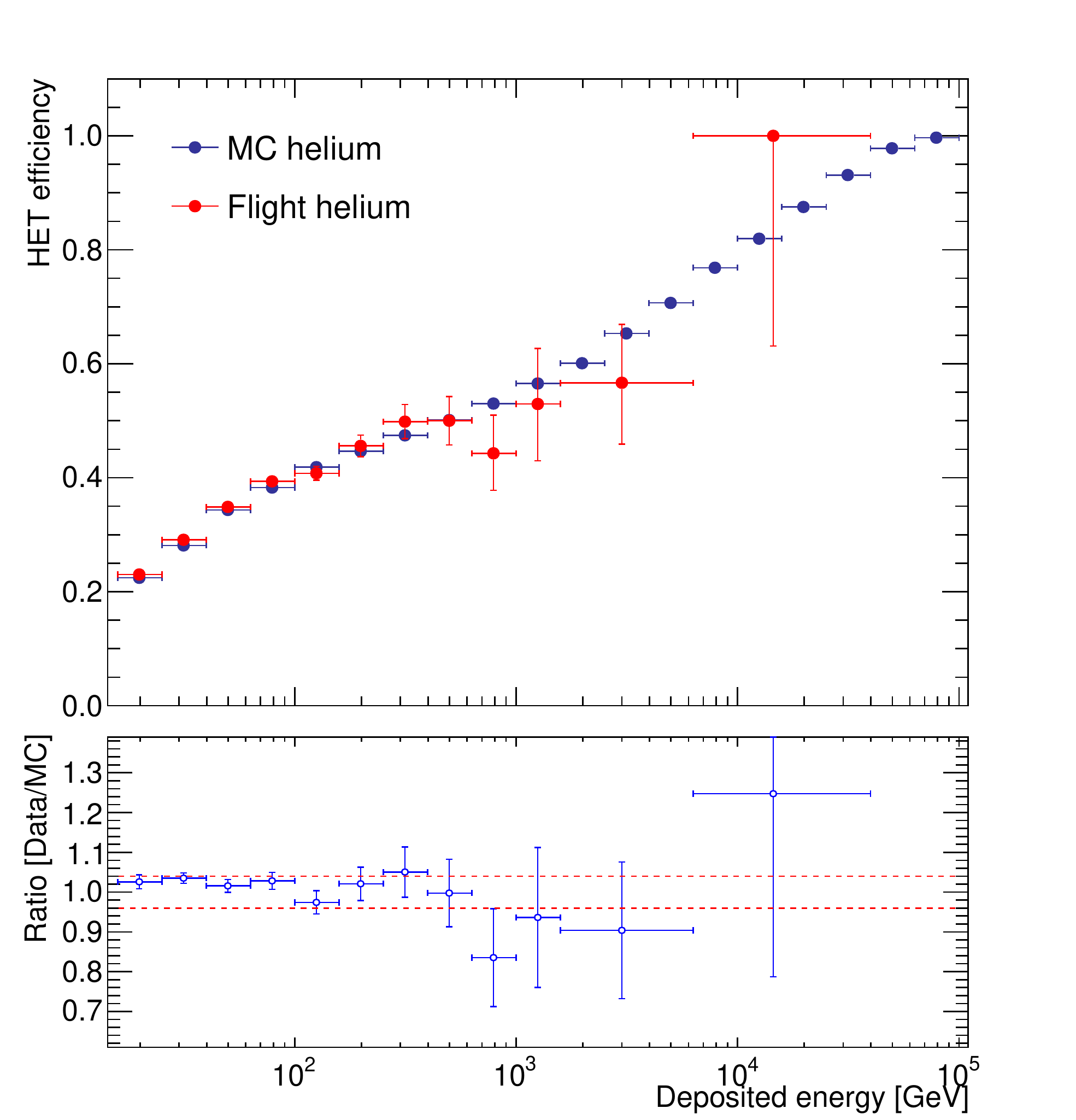}  
\caption{The HET efficiencies as functions of the BGO deposited 
energy for the selected helium candidates of the flight data (red) 
and MC simulations (dark blue). The differences between the flight 
data and MC simulations are about $4\%$ for energies up to $1$ TeV.}
\label{Fig:HET}
\end{figure}

The DAMPE detector implements four different triggers: the Unbiased trigger 
(UNBT), the Minimum Ionizing Particle trigger (MIPT), the Low Energy trigger 
(LET), and the HET [13]. Different pre-scale factors are applied for the 
UNBT, MIPT, and LET events, when the satellite operates at different 
latitudes. The UNBT, which is the least restrictive trigger, is used to 
estimate the HET efficiency. The HET efficiency is calculated as
\begin{equation}
\varepsilon_{\mathrm{HET}} = \frac{N_{\rm HET|UNBT}}{N_{\rm UNBT}},
\end{equation}
where $N_{\mathrm{HET|UNBT}}$ is the number of events with both the 
HET and UNBT activated, and $N_{\mathrm{UNBT}}$ is the number of 
the UNBT events. The resulting HET efficiencies as functions of the 
deposied energy inside the BGO calorimeter for both the flight and MC 
data are shown in Fig.~\ref{Fig:HET}. The UNBT sample has a pre-scale 
factor of 1/512 (1/2048) when the satellite operates in (out of) the 
geographical latitude range $[-20^\circ,20^\circ]$. Therefore at
high energies the statistical uncertainties of the flight data are
large. The relative differences between the flight data and the 
simulated data were estimated to be $\lesssim 4\%$, which were 
adopted as the systematic uncertainty from the trigger. 

\begin{figure}[!htb]
\includegraphics[width=0.43\textwidth]{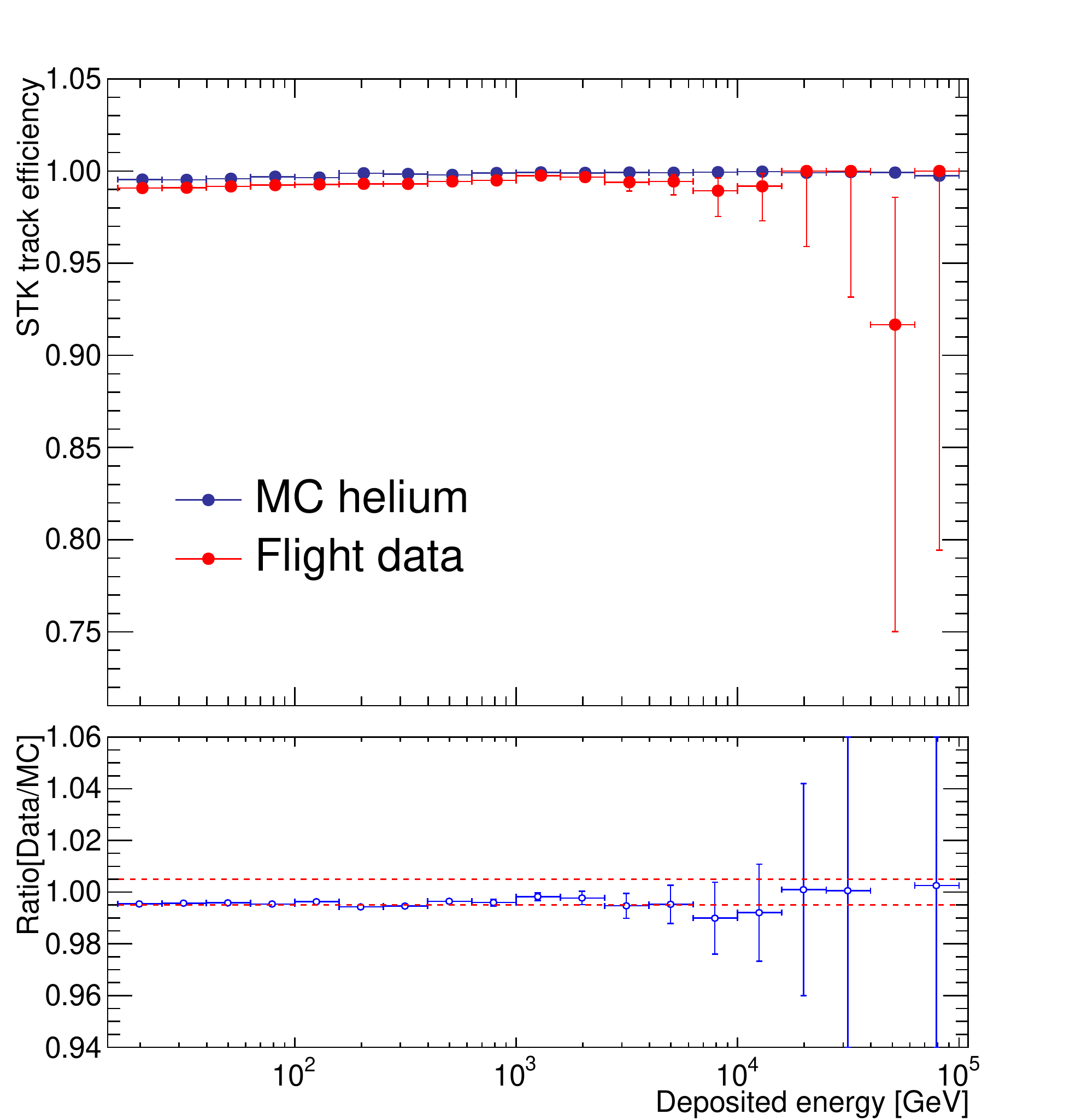} 
\caption{The STK track efficiencies as functions of the BGO deposited 
energy. The differences between the selected helium candidates of the 
flight data and MC simulations are about $0.5\%$.}
\label{Fig:Track}
\end{figure}

\begin{figure*}[!htb]
\centering
\includegraphics[width=0.45\textwidth]{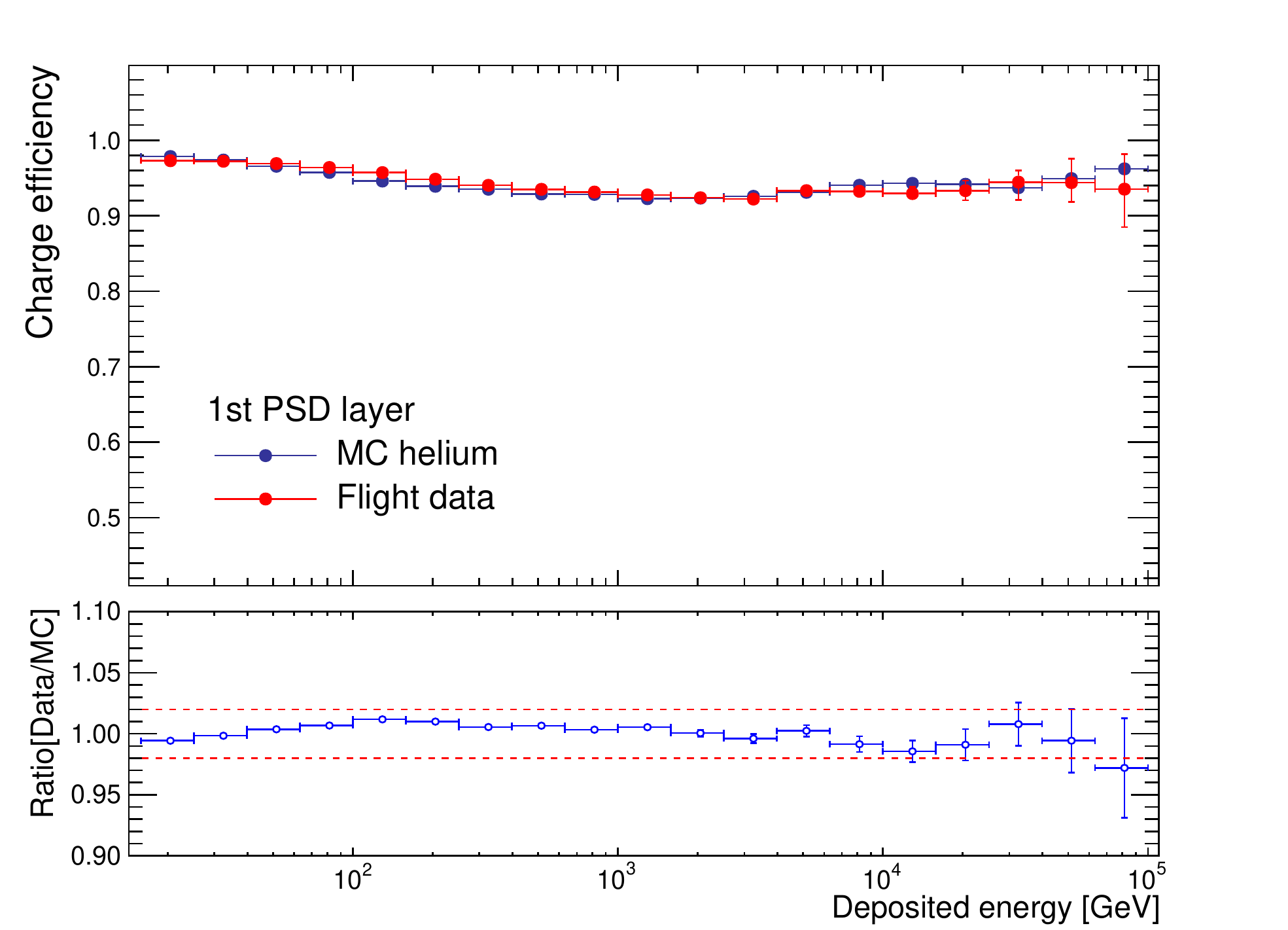}
\includegraphics[width=0.45\textwidth]{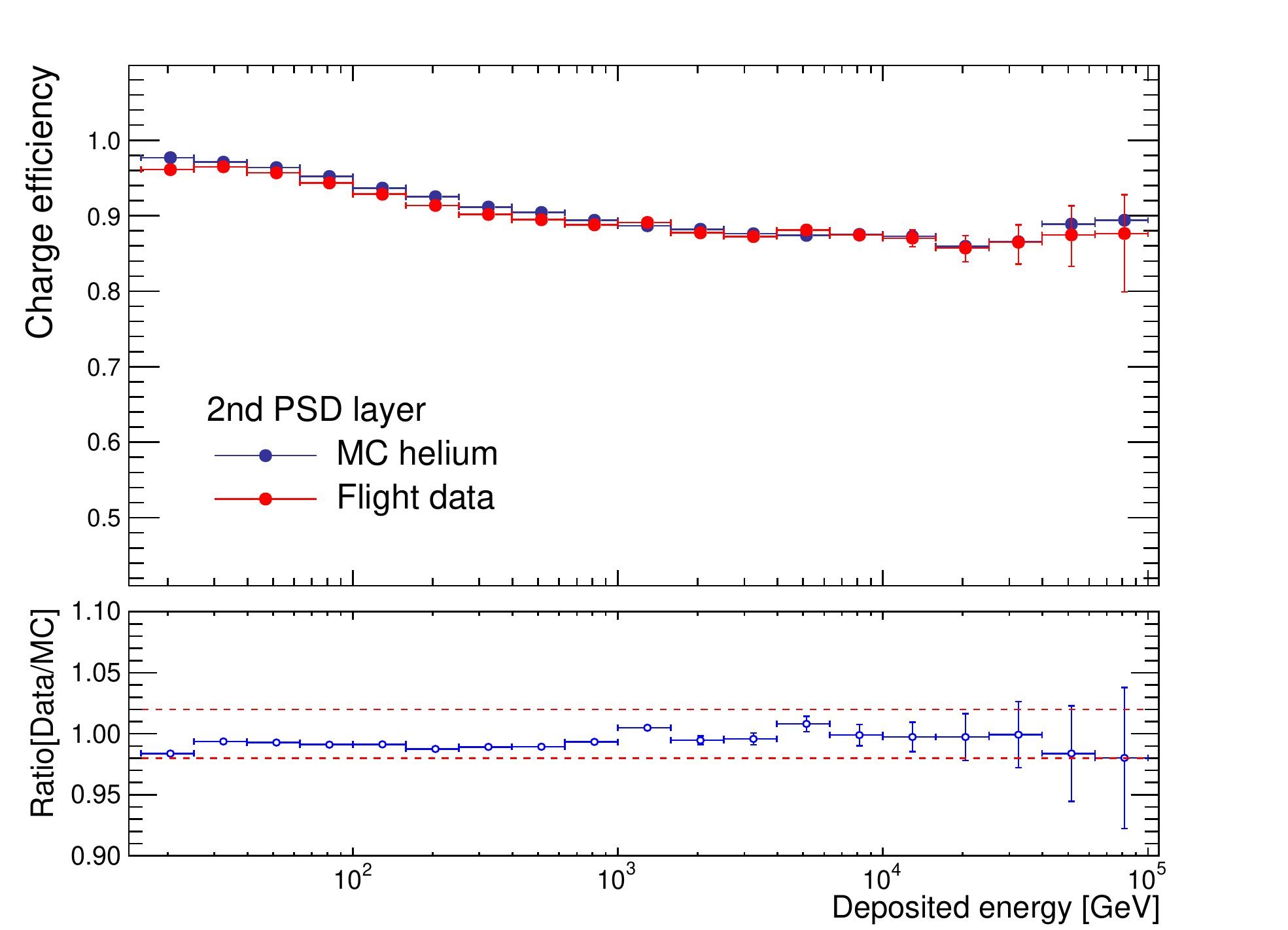}
\caption{Charge selection efficiencies for the two PSD layers as functions 
of the BGO deposited energy. The differences between the flight data and 
MC simulations are about $2\%$ for both PSD layers.}
\label{Fig:CHARGE}
\end{figure*}

\subsubsection{Track efficiency}

Given successful reconstruction of tracks of an event, two types of
tracks are available: the ones reconstructed by STK and the one by
BGO hits. The STK tracks are usually used in the standard analysis
since they are much more precise than the BGO track. The evaluation 
of the STK track efficiency (including both reconstruction and selection) 
is performed through choosing a helium sample based on BGO tracks and 
PSD charge, and then investigating the efficiency that passes the STK
track selection. The STK track efficiency is given by 
\begin{equation}
\varepsilon_{\rm track} = \frac{N_{\rm STK|BGO-PSD}}{N_{\rm BGO-PSD}},
\end{equation}
where $N_{\rm BGO-PSD}$ is the number of events selected with the BGO 
track matching with the PSD charge and $N_{\rm STK|BGO-PSD}$ is the
number of events which further passes the STK track selection used in
the present analysis. Fig.~\ref{Fig:Track} shows the comparison of
the track efficiency between the flight data and MC simulations, 
for different deposited energies. Small differences of about $0.5\%$ 
is observed.

\subsubsection{Charge selection efficiency}

The efficiencies related to the charge selection are estimated 
independently for each PSD layer, using the measurements provided by 
the first cluster point of the STK track. As an example, the efficiency 
of the first PSD layer is calculated as the ratio of event number selected 
using the charge of both PSD layers and the first cluster point of STK 
track ($N_{\rm PSD1|PSD2-STK1}$) to the number selected using only the
second PSD layer and the first cluster of STK track ($N_{\rm PSD2-STK1}$):
\begin{equation}
\varepsilon_{\rm PSD1} = \frac{N_{\rm PSD1|PSD2-STK1}}{N_{\rm PSD2-STK1}}.
\end{equation}
A similar way applies to the second PSD layer. 

\begin{figure}[!htb]
\centering
\includegraphics[width=0.45\textwidth]{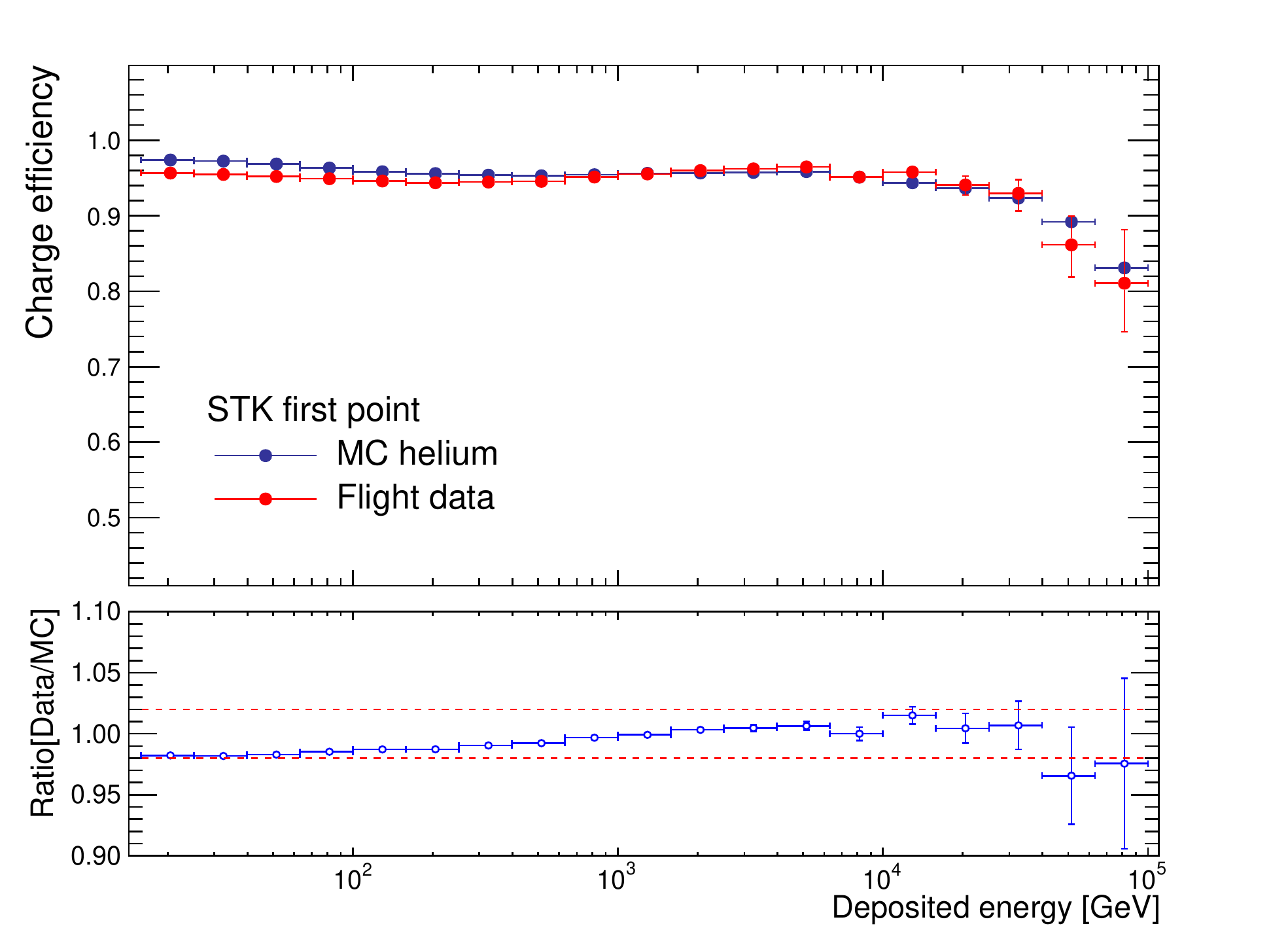}
\caption{Charge selection efficiencies for the first STK cluster point 
as functions of the BGO deposited energy. The differences between the
flight data and MC simulations are about $2\%$.}
\label{Fig:CHARGESTK}
\end{figure}

For the selection efficiency of the first STK cluster point, the control
sample is selected using the PSD. The STK charge efficiency is calculated as
\begin{equation}
\varepsilon_{\rm STK1} = \frac{N_{\rm STK1|PSD}}{N_{\rm PSD}}.
\end{equation}
The comparison between flight data and MC simulations are shown in 
Figs.~\ref{Fig:CHARGE} and \ref{Fig:CHARGESTK}. The differences are 
found to be about $2\%$ for both PSD layers and $2\%$ for the STK 
cluster point, respectively. The total systematic uncertainties from 
the charge selection are thus adopted as $3.5\%$.

\subsection{Background}
\begin{figure}[!htb]
\centering
\includegraphics[width=0.45\textwidth]{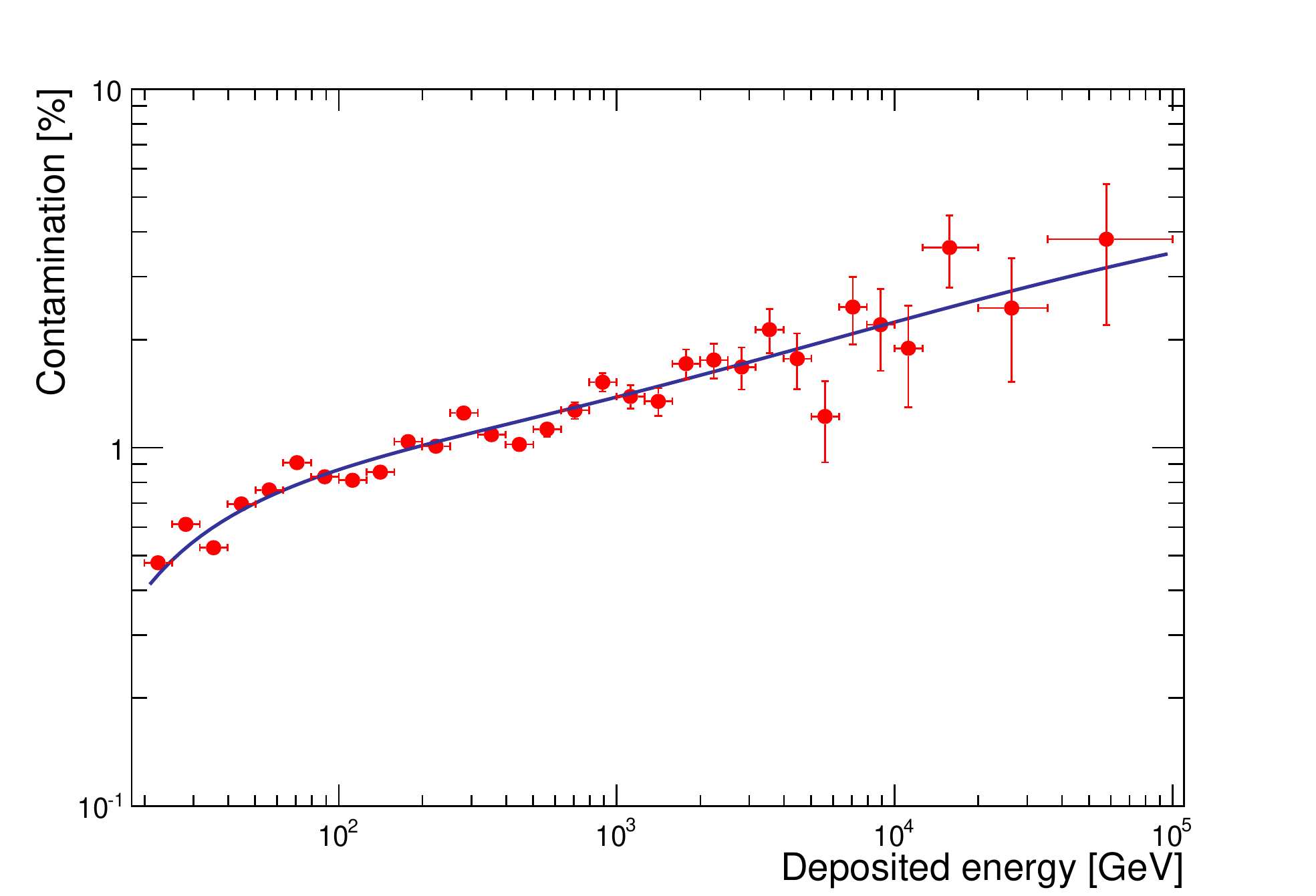}
\caption{Proton background percentage as a function of the BGO 
deposited energy. Solid line is a four-order polynomial fit, 
$\sum_{i=0}^{4}p_i\log^i(E_{\rm dep}/{\rm GeV})$.}
\label{Fig:Contamination}
\end{figure}

After our selection procedure, the remaining background of helium nuclei
is dominated by protons. The electron background and the background from
heavier nuclei (such as lithium) are negligibly small. The fit of the
PSD charge distribution (the minimum of the two PSD charge values) using 
the MC templates which are smeared to match with the data is performed 
in each deposited energy bin. The residual proton backgrounds are then 
estimated using the best-fit proton templates with the helium PSD charge 
selection of Eq.~(1). The results are presented in 
Fig.~\ref{Fig:Contamination}.

\subsection{Energy corrections}

The linear region of the energy measurement of a single BGO crystal can 
extend up to $\sim 4$ TeV for the dynode-2 readout device [13]. 
For very few highest energy events, saturation may occur for usually the 
BGO bar with the maximum energy deposition. A correction method based 
on MC simulations was developed based on the shower transverse and 
longitudinal developments [31], which was applied for the energy 
correction of those saturated events.

\begin{figure}[!htb]
\centering
\includegraphics[width=0.45\textwidth]{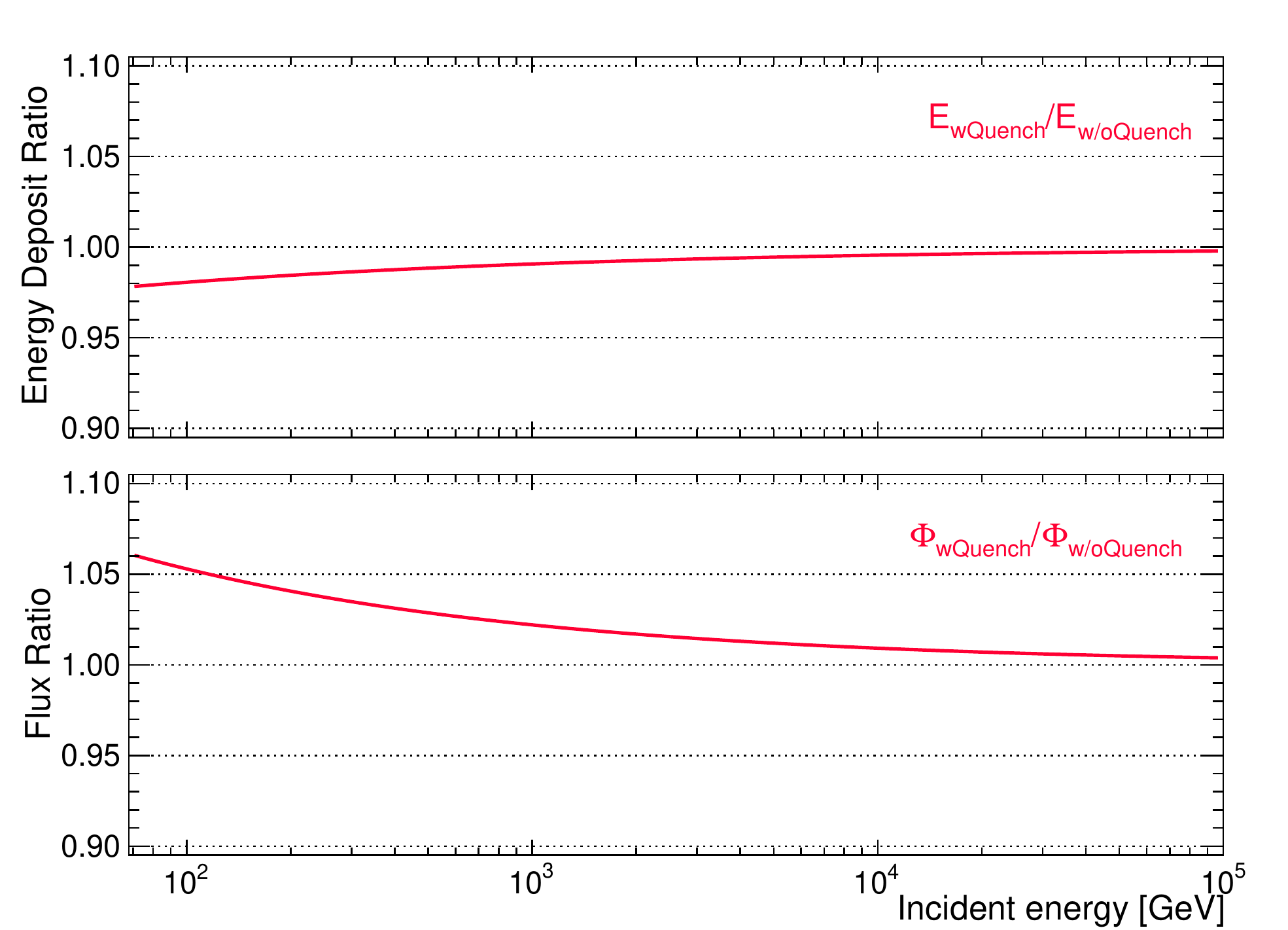}
\caption{The ratio of the quantities with the BGO quenching effect to 
that without the quenching effect. Top panel is for the mean energy 
deposition, and bottom panel is for the helium flux.}
\label{Fig:quenching}
\end{figure}

When the energies of secondary particles in the shower are low enough,
large amounts of ionization energy are deposited in the scintillator 
within very short traveling distances, resulting a nonlinearity between 
the scintillation photons and the ionization energy, known as the 
quenching effect [32]. The quenching effect would result in an 
under-estimate of the true energy of a shower. Using the test beam 
ion data and the ion MIP events from the flight data, the quenching
parameters of the DAMPE BGO scintillator were derived [33].
We implemented this quenching effect in the MC simulations, and 
investigated its impact on the energy measurement and response matrix 
calculation. The ratio of the mean energy deposition with the BGO 
quenching to that without the quenching is shown in the top panel of 
Fig.~\ref{Fig:quenching}. Considering the quenching effect will lead 
to $\sim2\%$ ($0.2\%$) lower energy deposition for helium incident 
energy of 80 GeV (80 TeV). Using the corresponding response matrix, 
we get the helium spectrum, whose ratio to the spectrum without 
considering the quenching effect is shown in the bottom panel of 
Fig.~\ref{Fig:quenching}. The impact on the unfolded spectrum varies 
from $\sim 5.5\%$ at 80 GeV to $\sim 0.4\%$ at 80 TeV.

\begin{figure}[!htb]
\centering
\includegraphics[width=0.45\textwidth]{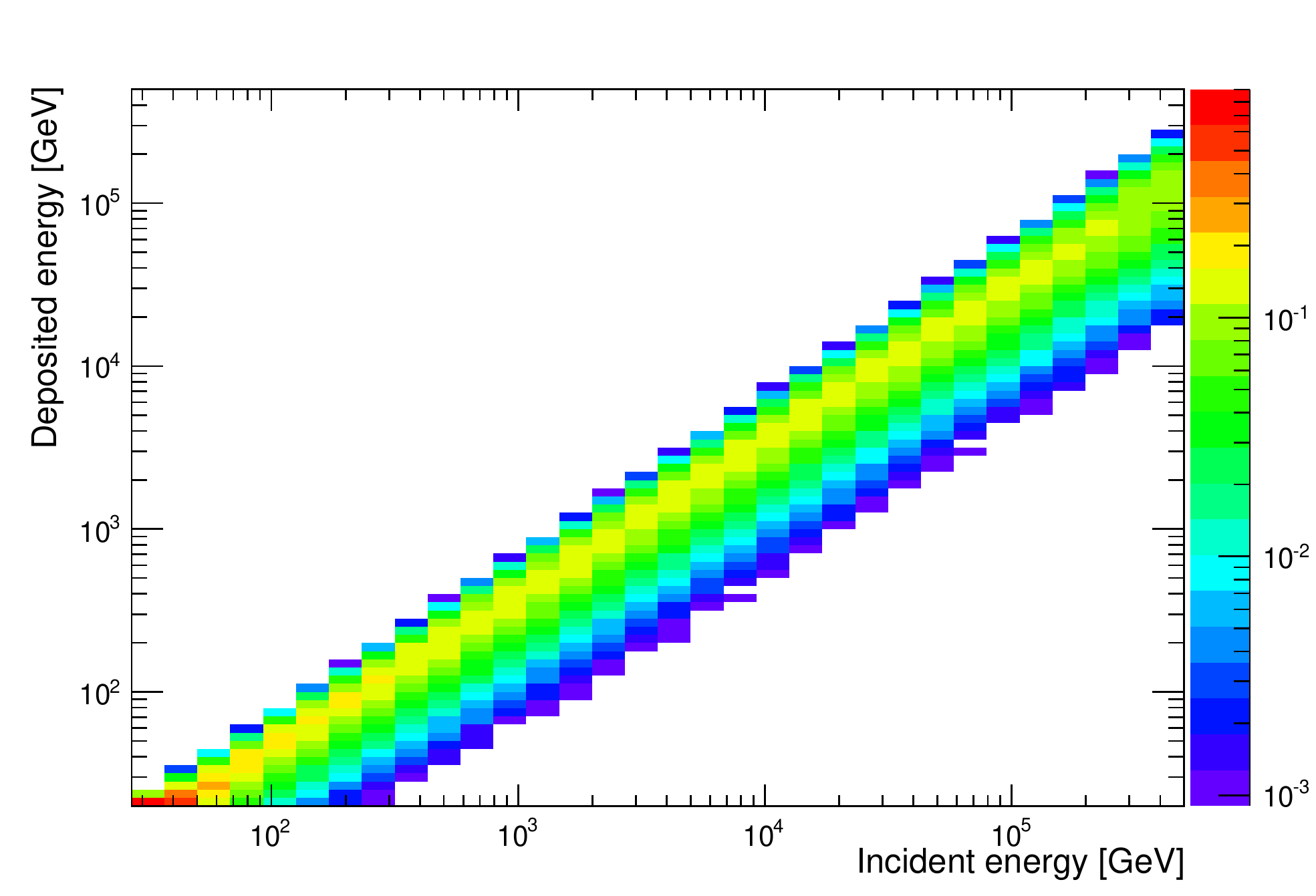}
\caption{Response matrix used in the unfolding procedure obtained from
the selected MC helium sample.}
\label{Fig:matrix}
\end{figure}

Fig.~\ref{Fig:matrix} shows the energy response matrix based on the
GEANT4 FTFP\_BERT model, after including the quenching effect.
The color represents the relative probability that a helium nucleus
with incident energy $E_{\rm inc}$ deposits $E_{\rm dep}$ energy in the
calorimeter. The energy resolution of helium can thus be inferred to 
be about $25\%\sim35\%$ for incident energies from 100 GeV to 80 TeV.

\subsection{Observed counts, unfolded fluxes, and systematic uncertainties}

Table~\ref{tab:counts} gives the numbers of selected helium candidates 
for deposited energies from 20 GeV to 32 TeV. The contamination from
protons as shown in Fig.~\ref{Fig:Contamination} has not been subtracted.

\begin{table}[ht]
\begin{center}
\caption{Measured numbers of helium candidates from the flight data.}
\begin{tabular}{ccc} \hline\hline
$E_{\rm dep}^{\rm min}$  &  $E_{\rm dep}^{\rm max}$   &   Counts    \\ 
($10^3$~GeV) & ($10^3$~GeV) & \\  \hline
0.020 & 0.025 &	3921850 \\
0.025 &	0.032 &	3379910 \\
0.032 &	0.040 &	2748360 \\
0.040 &	0.050 &	2128160 \\
0.050 &	0.063 &	1586940 \\
0.063 &	0.079 &	1145530 \\
0.079 &	0.100 &	808389 \\
0.100 &	0.126 &	562406 \\
0.126 &	0.158 &	386395 \\
0.158 &	0.200 &	264417 \\
0.200 &	0.251 &	180063 \\
0.251 &	0.316 &	122885 \\
0.316 &	0.398 &	83617 \\
0.398 &	0.501 &	57103 \\
0.501 &	0.631 &	39222 \\
0.631 &	0.794 &	27143 \\
0.794 &	1.000 &	19057 \\
1.000 &	1.259 &	13463 \\
1.259 &	1.585 &	9538 \\
1.585 &	1.995 &	6664 \\
1.995 &	2.512 &	4747 \\
2.512 &	3.162 &	3344 \\
3.162 &	3.981 &	2486 \\
3.981 &	5.012 &	1871 \\
5.012 &	6.310 &	1319 \\
6.310 &	7.943 &	950 \\
7.943 &	10.00 &	720 \\
10.00 &	12.59 &	561 \\
12.59 &	15.85 &	332 \\
15.85 &	19.95 &	243 \\
19.95 &	25.12 &	157 \\
25.12 &	31.62 &	99 \\ 
\hline
\end{tabular}
\label{tab:counts}
\end{center}
\end{table}

\begin{figure}[!htb]
\centering
\includegraphics[scale=0.45]{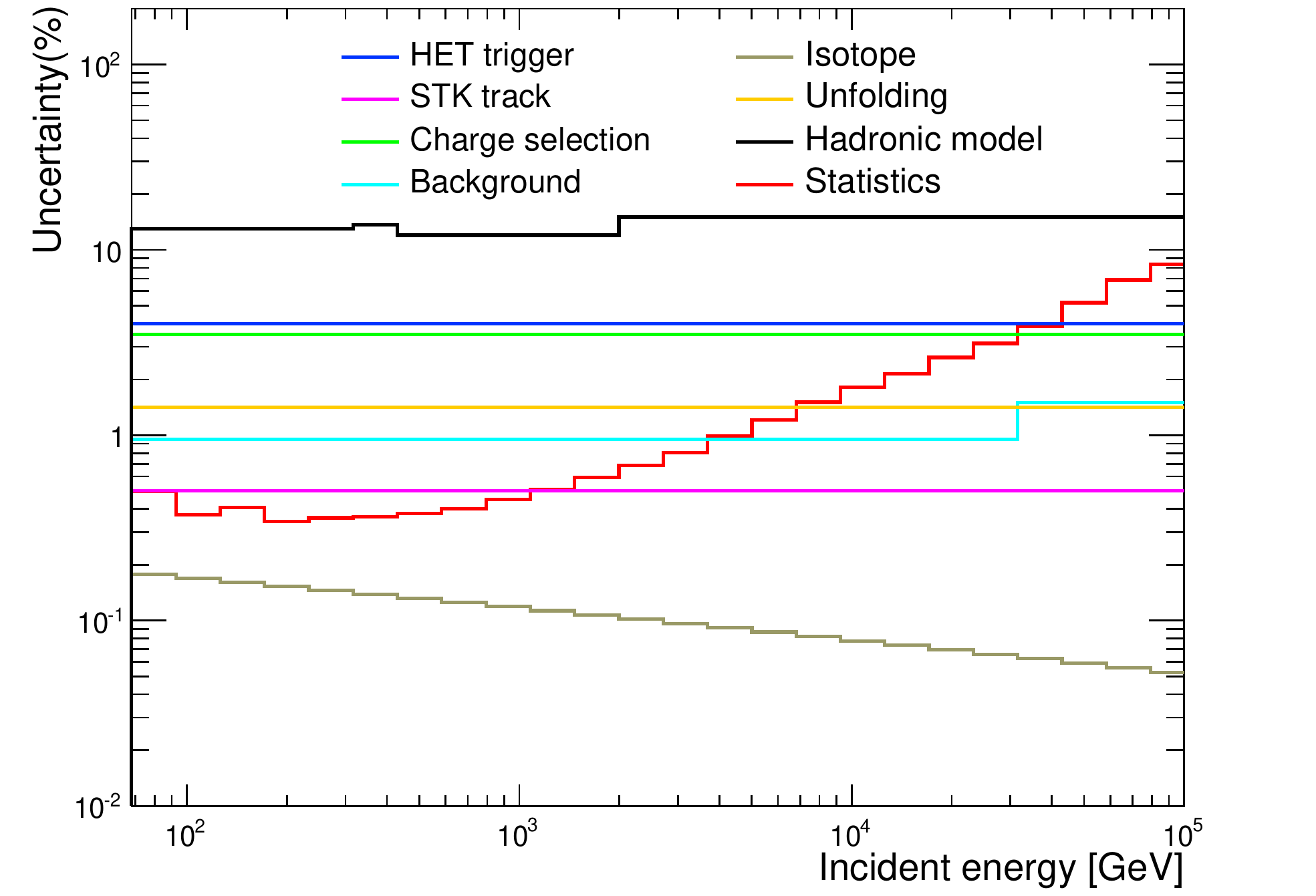}
\caption{Relative statistical and systematic uncertainties of the measured
helium spectrum.}
\label{Fig:systematics}
\end{figure}

Table \ref{tab:flux} gives the helium fluxes after the background 
subtraction and the unfolding procedure. The relative uncertainties
of the fluxes are shown in Fig.~\ref{Fig:systematics}.

\begin{table}[ht]
\centering
\footnotesize
\caption{Fluxes of helium nuclei measured with DAMPE, together with the
$1\sigma$ statistical uncertainties ($\sigma_{\rm stat}$) and the
systematic uncertainties from the analysis ($\sigma_{\rm ana}$) and
from the hadronic interaction models ($\sigma_{\rm had}$).}
\begin{tabular}{ccccccccccc}
\toprule
\hline
\hline
$E_{\rm min}$ & $E_{\rm max}$  & $\langle E \rangle$ & $\Phi \pm \sigma_{\rm stat} \pm \sigma_{\rm ana} \pm \sigma_{\rm had}$ \\
($10^3$ GeV) & ($10^3$ GeV)  & ($10^3$ GeV) & [GeV$^{-1}$~m$^{-2}$~s$^{-1}$~sr$^{-1}$]
 \\
\hline
\midrule
0.068 & 0.093 & 0.079 & ($5.261 \pm 0.025 \pm 0.295 \pm 0.684$) $\times 10^{-2}$\\
0.093 & 0.125 & 0.108 & ($2.369 \pm 0.010 \pm 0.133 \pm 0.308$) $\times 10^{-2}$\\
0.125 & 0.171 & 0.146 & ($1.066 \pm 0.004 \pm 0.060 \pm 0.139$) $\times 10^{-2}$\\
0.171 & 0.232 & 0.199 & ($4.658 \pm 0.015 \pm 0.261 \pm 0.606$) $\times 10^{-3}$\\
0.232 & 0.316 & 0.270 & ($2.022 \pm 0.007 \pm 0.113 \pm 0.263$) $\times 10^{-3}$\\
0.316 & 0.430 & 0.367 & ($8.846 \pm 0.031 \pm 0.495 \pm 1.207$) $\times 10^{-4}$\\
0.430 & 0.584 & 0.499 & ($3.894 \pm 0.014 \pm 0.218 \pm 0.467$) $\times 10^{-4}$\\
0.584 & 0.794 & 0.679 & ($1.727 \pm 0.007 \pm 0.097 \pm 0.207$) $\times 10^{-4}$\\
0.794 & 1.079 & 0.923 & ($7.674 \pm 0.034 \pm 0.430 \pm 0.921$) $\times 10^{-5}$\\
1.079 & 1.466 & 1.254 & ($3.422 \pm 0.017 \pm 0.192 \pm 0.411$) $\times 10^{-5}$\\
1.466 & 1.993 & 1.705 & ($1.549 \pm 0.009 \pm 0.087 \pm 0.186$) $\times 10^{-5}$\\
1.993 & 2.709 & 2.317 & ($7.161 \pm 0.048 \pm 0.401 \pm 1.074$) $\times 10^{-6}$ \\
2.709 & 3.682 & 3.150 & ($3.348 \pm 0.027 \pm 0.187 \pm 0.502$) $\times 10^{-6}$\\
3.682 & 5.005 & 4.281 & ($1.548 \pm 0.015 \pm 0.087 \pm 0.232$) $\times 10^{-6}$\\
5.005 & 6.803 & 5.819 & ($7.021 \pm 0.084 \pm 0.393 \pm 1.053$) $\times 10^{-7}$\\
6.803 & 9.247 & 7.910 & ($3.227 \pm 0.048 \pm 0.181 \pm 0.484$) $\times 10^{-7}$\\
9.247 & 12.57 & 10.75 & ($1.552 \pm 0.028 \pm 0.087 \pm 0.233$) $\times 10^{-7}$\\
12.57 & 17.08 & 14.61 & ($7.451 \pm 0.158 \pm 0.417 \pm 1.118$) $\times 10^{-8}$\\
17.08 & 23.22 & 19.86 & ($3.504 \pm 0.091 \pm 0.196 \pm 0.526$) $\times 10^{-8}$\\
23.22 & 31.56 & 27.00 & ($1.654 \pm 0.051 \pm 0.093 \pm 0.248$) $\times 10^{-8}$\\
31.56 & 42.90 & 36.70 & ($7.416 \pm 0.286 \pm 0.415 \pm 1.112$) $\times 10^{-9}$\\
42.90 & 58.32 & 49.88 & ($3.104 \pm 0.160 \pm 0.174 \pm 0.466$) $\times 10^{-9}$\\
58.32 & 79.27 & 67.80 & ($1.299 \pm 0.089 \pm 0.073 \pm 0.195$) $\times 10^{-9}$\\

\hline
\bottomrule
\end{tabular}
\label{tab:flux}
\end{table}

\subsection{Spectral fitting}

To quantify the spectral features, the fits to the helium spectrum
(Table~\ref{tab:fit}) are made. The function used in the fit is a 
smoothly broken power-law (SBPL) form
\begin{equation}
\Phi(E) = \Phi_0 \left( \frac{E}{\rm TeV} \right)^{-\gamma}
\left[1 + \left(\frac{E}{E_B} \right)^s \right]^{\Delta\gamma/s},
\label{eq:SBPL}
\end{equation}
where $\Phi_0$ is the flux normalization, $\gamma$ is the spectral
index for energies far below the break energy $E_B$, $\Delta\gamma$ is
the change of the spectral indices above $E_B$, and $s$ is a parameter
describing the smoothness of the break. 

The helium spectrum is characterized by a clear hardening at $\sim$TeV 
followed by a softening approaching $\sim30$~TeV (Fig.~3). We therefore 
carry out the fit in different energy ranges to address these features 
individually. To take into account possible bin-to-bin correlations of 
the systematic uncertainties, we employ the nuisance parameter method as 
described in detail in Refs.~[7, 36]. The $\chi^2$ function is thus defined as
\begin{eqnarray}
\chi^2&=&\sum_{i=k}^{n}\sum_{j=k}^{n}[\Phi(E_{i})S(E_{i};\,\boldsymbol{w})
-\Phi_i]{\mathcal C}^{-1}_{ij}[\Phi(E_{j})S(E_{j};\,\boldsymbol{w})-\Phi_j] 
\nonumber \\
& + & \sum_{\ell=1}^{m} \left(\frac{1-w_\ell}
{\tilde{\sigma}_{\rm sys,\ell}}\right)^2,
\end{eqnarray}
where $E_i$ and $\Phi_i$ are the median energy and flux of the measurement 
in the $i$-th energy bin, ${\mathcal C}$ is the covariance matrix of the
fluxes derived from the toy MC simulation when evaluating the statistical
uncertainties, $\Phi(E_{i})$ is the model predicted flux, 
$S(E_{i};\,\boldsymbol{w})$ is a piecewise function defined by its value 
$\boldsymbol{w}$, and $\tilde{\sigma}_{\rm sys,\ell}=\sigma_{\rm ana}/\Phi$
is the relative systematic uncertainty of the data in corresponding energy 
range covered by the $\ell$-th nuisance parameter. The nuisance parameters 
enable flux adjustments in various bins. Note that here we single out
the systematic uncertainties from the hadronic models.

\begin{table}[ht]
\begin{center}
\caption{Parameters from the fits with SBPL in two different energy
ranges of the Helium spectrum.}
\begin{tabular}{lll} \hline\hline
& Hardening  & Softening  \\ 
Fit range & $[0.32-5.0]$~TeV  & $[6.8-80]$~TeV \\
Nuisance parameters  & 3  &  2            \\ \hline
$\Phi_0$~($10^{-5}$~GeV$^{-1}$~m$^{-2}$~s$^{-1}$~sr$^{-1}$) & $6.08^{+0.22+0.00}_{-0.25-0.64}$  & $4.71^{+0.27+0.00}_{-0.25-0.56}$  \\
$\gamma$  & $2.68^{+0.02+0.00}_{-0.01-0.05}$  & $2.41^{+0.02+0.02}_{-0.02-0.00}$ \\
$E_B$~(TeV) & $1.25^{+0.15+1.05}_{-0.12-0.00}$   & $34.4^{+6.7+11.6}_{-9.8-0.0}$ \\
$\Delta\gamma$ & $0.18^{+0.05+0.00}_{-0.02-0.06}$ & $-0.51^{+0.18+0.01}_{-0.20-0.00}$ \\
$s$ & $3.6^{+2.3+13.4}_{-1.6-0.0}$ & 5.0 (fixed) \\ \hline\hline
\end{tabular}
\label{tab:fit}
\end{center}
\end{table}

\begin{figure}[!htb]
\centering
\includegraphics[width=0.45\textwidth]{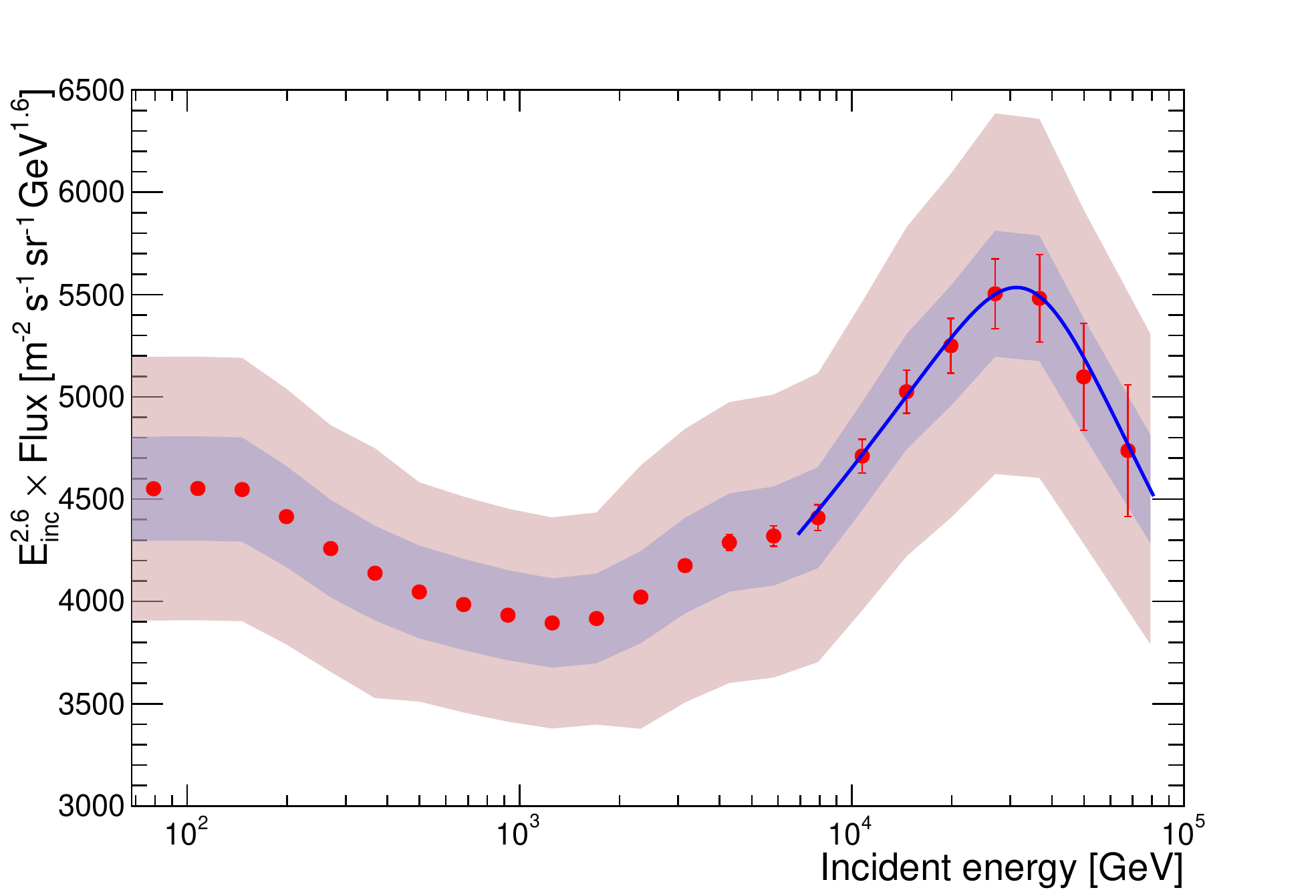}
\caption{Best-fit of the helium flux with a SBPL function (blue line) 
in the energy range $[6.8-80]$~TeV, compared with the data. Error bars 
correspond to the statistical uncertainties, and the shaded bands show 
the systematic uncertainties.} 
\label{fig:fit}
\end{figure}

For the hardening feature, the fit is performed in the range of $[0.32-5.0]$ 
TeV, and 3 nuisance parameters are adopted. The SBPL fit results in a 
reduced chi-squared value of $\chi^2/{\rm dof}=4.7/1$, where dof is the 
number of degrees of freedom. The $\chi^2$ value is big, possibly due 
to the small irregularities of the spectrum induced by the unfolding. 
The parameters are given in Table~\ref{tab:fit}. Compared with a single 
power-law model fit with $\chi^2/{\rm dof}=619.8/4$, the SBPL model is 
favored at a significance of $\sim 24.6\sigma$ for three more free 
parameters. A caveat is that this significance may be over-estimated 
given the large $\chi^2$ values in both fits with the SBPL and single 
power-law models. The spectral indices of the PAMELA helium measurement 
are $\gamma_1=2.766\pm0.029$, $\gamma_2=2.477\pm0.067$ for rigidities 
below and above $243^{+27}_{-31}$~GV [4]. The results of AMS-02 are 
$\gamma=2.780\pm0.007$, $\Delta\gamma=0.119\pm0.033$, and the break 
rigidity is $245\pm46$~GV [6]. The low-energy spectral index of the 
DAMPE measurement
($2.68^{+0.02}_{-0.01}$) is slightly harder than those of PAMELA 
and AMS-02. The value of $\Delta\gamma$ of the DAMPE measurement 
($0.18^{+0.05}_{-0.02}$) lies between those of PAMELA and AMS-02. 
The break energy we get ($1.25^{+0.15}_{-0.12}$~TeV) is higher than 
those of PAMELA and AMS-02. Part of the differences may come from 
the different energy ranges adopted in the fits. 
To estimate the effect on the fitting parameters from the hadronic models, 
we carry out separate fit to the fluxes derived with the FLUKA simulations, 
and the differences are given as the second errors in Table~\ref{tab:fit}.

The softening is studied in the range of $[6.8 - 80]$~TeV. 
We adopt 2 nuisance parameters in this narrow energy range.
Given the relatively large uncertainties of the data, the smoothness
parameter $s$ cannot be effectively constrained by the data, and
we fix it to be 5 for a consistency with that adopted in our proton
analysis paper [7]. The fitting results are 
$\Phi_0=4.71^{+0.27}_{-0.25}\times10^{-5}$ 
GeV$^{-1}$~m$^{-2}$~s$^{-1}$~sr$^{-1}$, $\gamma=2.41^{+0.02}_{-0.02}$, 
$\Delta\gamma=-0.51^{+0.18}_{-0.20}$, $E_B=34.4^{+6.7}_{-9.8}$ TeV,
and $\chi^2/{\rm dof}=2.53/2$. For a fit with a single power-law
function we get $\chi^2/{\rm dof}=24.25/4$. Therefore we get a
significance of the spectral softening of $\sim4.3\sigma$, given two 
more free parameters of the SBPL model. The best-fitting result of the 
softening structure, together with the DAMPE measurements, is shown in 
Fig.~\ref{fig:fit}. Compared with the $13.6$ TeV break energy of the 
DAMPE proton spectrum [7], the softening energies of both protons 
and helium nuclei are consistent with a charge-dependent scenario. 
The break energy is higher if the FLUKA simulation is used
(see Table~\ref{tab:fit}). Therefore our current results cannot 
rule out a mass-dependent softening scenario.

We also assume an exponentially cutoff power-law (ECPL) function 
to describe the softening feature, and the fitting gives 
$\Phi_0=4.19^{+0.31}_{-0.29}\times10^{-5}$
GeV$^{-1}$~m$^{-2}$~s$^{-1}$~sr$^{-1}$, $\gamma=2.31^{+0.04}_{-0.04}$,
$E_{\rm cut}=117.6^{+35.8}_{-22.8}$ TeV, and $\chi^2/{\rm dof}=4.15/3$.
The current data may not be able to distinguish the ECPL model from
the SBPL one. We expect that future measurement of the helium spectrum
to higher energies by DAMPE with larger statistics and better control 
of systematic uncertainties will be very helpful in testing the 
detailed behavior of the spectral softening.


\end{document}